\documentclass[a4paper,11pt]{article}

\usepackage{jheppub} 
\usepackage{etoolbox}
\patchcmd{\maketitle}{\@fpheader}{}{}{}
\usepackage[T1]{fontenc} 
\usepackage[utf8]{inputenc} 
\usepackage{microtype} 
\usepackage{lmodern} 
\usepackage{mdframed} 
\usepackage{mathtools}
\usepackage{graphicx}
\usepackage{cancel} 
\usepackage{todonotes} 
\usepackage[normalem]{ulem}
\usepackage{soul}
\usepackage{amsmath}
\usepackage{amsfonts}
\usepackage{amssymb}
\usepackage{stmaryrd}
\usepackage[mathscr]{euscript}
\usepackage{mathtools}
\usepackage{mathrsfs}
\usepackage{multicol}
\usepackage{mleftright} \mleftright
\usepackage{tensor} 
\usepackage{braket} 
\usepackage{mathrsfs} 
\usepackage{bbold}
\usepackage{color}
\usepackage{braket}
\usepackage{slashed}
\usepackage{comment}
\usepackage{physics}
\usepackage{float}
\usepackage{enumitem}
\usepackage{tikz}
\usetikzlibrary{positioning,shapes.geometric,calc,arrows.meta}

\providecommand{\sc}{{ }}
\renewcommand{\sc}{{ }}

\DeclareMathAlphabet{\mathfs}{U}{rsfs}{m}{n}                     %

\newcommand{\be}{\nopagebreak[3]\begin{equation}}
\newcommand{\ee}{\end{equation}}
\newcommand{\bee}{\nopagebreak[3]\begin{equation*}}
\newcommand{\eee}{\end{equation*}}
\newcommand{\ba}{\nopagebreak[3]\begin{eqnarray}}
\newcommand{\ea}{\end{eqnarray}}
\newcommand{\baa}{\nopagebreak[3]\begin{eqnarray*}}
\newcommand{\eaa}{\end{eqnarray*}}
\newcommand{\bal}{\nopagebreak[3]\begin{aligned}}
\newcommand{\eal}{\end{aligned}}

\newcommand{\bseq}{\nopagebreak[3]\begin{subequations}}
\newcommand{\eseq}{\end{subequations}\noindent}

\newcommand{\mH}{\mathcal{H}}

\newcommand{\mZ}{\mathcal{Z}}

\title{Quantum Fluctuations and Newton-Cartan Geometry for Non-Relativistic de Sitter space}

\author[1]{Matthias Harksen}
\author[1,2]{\hspace{-0.1cm}, Diego Hidalgo}
\author[3]{\hspace{-0.1cm}, and Watse Sybesma}
\affiliation[1]{Science Institute, University of Iceland,\\ Dunhaga 3, 107 Reykjav\'ik, Iceland}
\affiliation[2]{Institute for Theoretical Physics, Utrecht University,\\
Princetonplein 5, 3584 CE Utrecht, The Netherlands}
\affiliation[3]{Nordita, KTH Royal Institute of Technology and Stockholm University \\ Hannes Alfvéns v\"ag 12, SE-106 91 Stockholm, Sweden}
\emailAdd{mhb6@hi.is}
\emailAdd{dhidalgo@hi.is}
\emailAdd{watse.sybesma@su.se}
\preprint{{\bf } }

\abstract{
We study a non-relativistic realisation of two-dimensional de Sitter gravity both from its boundary and bulk description with the goal of learning about de Sitter space and paving the way for extending the holographic duality into a non-relativistic direction.
On the boundary side, we analyse the Schwarzian-type boundary action associated with non-relativistic de Sitter gravity and evaluate its one-loop partition function in order to compute its quantum fluctuations. Rather than relying on the coadjoint-orbit construction, we derive the path integral measure directly from the action using the Ostrogradsky formalism. We find a temperature-dependent prefactor scaling as $T^2$, of which the power agrees with the counting of the four global symmetry generators present.
On the bulk side, we construct the corresponding torsionless Newton-Cartan geometry and show that it satisfies the equations of motion of a non-relativistic JT-like action and uplift the geometry to a three-dimensional Lorentzian geometry. 
} 

\begin{document}
\maketitle

\section{Introduction}
It is difficult to overstate the implications of the initial discovery of the Anti-de Sitter/conformal field theory (AdS/CFT) duality on our understanding of physics, with a central role being played by $\mathcal{N}=4$ Super Yang-Mills, on one side of the duality, and type II-B supergravity on an AdS background on the other side \cite{Maldacena:1997re}. More recently, our understanding of the duality has been significantly extended in the realm of two-dimensional gravity, which can be considered the lowest-dimensional realisation of the AdS/CFT correspondence and, in this manner, facilitates the study of quantum gravity due to its technical manageability. In this context, the bulk gravity model containing an AdS background is governed by Jackiw-Teitelboim (JT) gravity theory \cite{Teitelboim:1983ux,Jackiw:1984je}, where the so-called near-AdS$_2$ dynamics is controlled by a boundary action involving the Schwarzian derivative. It was found that he same action also governs the low-energy regime of the $(0+1)$-dimensional Sachdev-Ye-Kitaev (SYK) model \cite{Sachdev:1992fk, KitaevTalks, Sachdev:2010um, Maldacena:2016hyu, Sarosi:2017ykf, Gu:2019jub, Mertens:2018fds, Grumiller:2002nm}. Some features of the SYK model, such as the infrared ``nearly'' conformal invariance and chaotic behaviour, see e.g. \cite{Sachdev:1992fk}, suggest the possible existence of a gravitational dual model in AdS$_2$ spacetime.

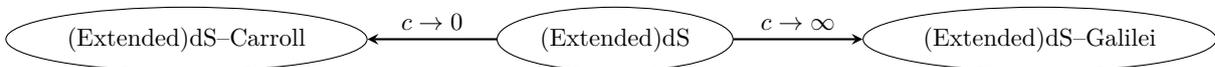
\begin{figure}[ht!]
    \centering
    \vspace{0.25cm}
    \begin{tikzpicture}[
        scale=0.85,
        transform shape,
        >=stealth,
        node distance=2cm,
        kin/.style={draw, ellipse, minimum width=2.8cm, minimum height=1cm}
    ]

        \node[kin] (ads) {$(\mathrm{Extended})\mathrm{dS}$};
        \node[kin, left=of ads]  (carroll) {$(\mathrm{Extended})\mathrm{dS}$--Carroll};
        \node[kin, right=of ads] (galilei) {$(\mathrm{Extended})\mathrm{dS}$--Galilei};

        \draw[->, thick] (ads) -- (galilei)
            node[midway, above] {$c \to \infty$};

        \draw[->, thick] (ads) -- (carroll)
            node[midway, above] {$c \to 0$};

    \end{tikzpicture}
    \caption{The figure shows two kinematical contractions of de Sitter (dS) symmetry algebra. Here $c$ denotes the contraction parameter, which may be interpreted heuristically as the speed of light. The extended version refers to the possibility of adding Abelian generators to the dS algebra, thereby yielding extended versions of the non-Lorentzian algebras.  }
    \label{fig:ads-cross}
\end{figure}

Amid the intense efforts to extend the AdS/CFT duality to a more general notions of holography that also encompasses different spacetimes, most of that effort has focused on ensuring the {\it relativistic} symmetries in the bulk. Going beyond this relativistic bulk context, there has been an emergence of explorations of non-Lorentzian geometries \cite{Christensen:2013lma, Hartong:2015xda, Hansen:2020pqs, Hansen:2021fxi,  Bergshoeff:2022eog} and non-Lorentzian string theories \cite{Gomis:2000bd, Danielsson:2000gi, Cardona:2016ytk, Bidussi:2021ujm, Oling:2022fft, Harksen:2024bnh}. These research topics, initially fuelled by the desire to learn about non-relativistic quantum field theories in the context of condensed matter, see e.g. \cite{Taylor:2015glc}, grew into a field of its own graduating beyond its original scope. The term non-Lorentzian here is used as an umbrella term for both the non-relativistic (``speed of light to infinity'') and Carrollian (``speed of light to zero'') limits (see Figure \ref{fig:ads-cross}). Particularly, the non-relativistic limit can be complemented with the cosmological constant vanishing $\Lambda \to 0$, so that the dS or AdS algebra becomes the Newton-Hooke algebra \cite{Leblonde:1965, Brugues:2006yd, Derome1972HookesSA}, the analogue of the Galilei algebra in the presence of a universal cosmological repulsion or attraction constant, respectively. The idea of considering the non-relativistic version of de Sitter space was explored for its possible application to non-relativistic cosmology in \cite{Leblonde:1965, Aldrovandi:1998im} and it was also considered in the context of M- or String theory where a modification of the matrix model is obtained using the corresponding symmetry algebra \cite{Gao:2001sr}. In the present paper, we will utilise an centrally extended version of the so-called Newton-Hooke symmetry algebra that can be obtained with a central generator. This central generator can be thought of as the analogue of the Bargmann generator, which allows for the representation of `massive' states. In its two-dimensional incarnation will refer to it as the Extended de Sitter Galilean (EdS-G) symmetry algebra following \cite{Grumiller:2020elf}. 

The present paper is devoted to investigating the EdS-G gravity, which allows us to address holography in a controlled non-relativistic de Sitter setting as well as learning about de Sitter space. We will explore both the boundary and bulk realisation of this non-relativistic version of de Sitter gravity.

On the boundary side we will use the BF formalism in two dimensions, a first-order gauge theory formulation of a gravity theory that realises EdS-G isometries, for which a boundary formulation was derived in \cite{Gomis:2020wxp} -- see also \cite{Grumiller:2020elf} for studies of non-Lorentzian realisations of JT. It was shown that the asymptotic structure of this Galilean gravity is governed by a so-called non-relativistic Schwarzian theory, which exhibits an infinite-dimensional symmetry algebra given by the twisted Virasoro algebra. 
Inspired by the computation of the one-loop contribution to the boundary description of the relativistic Schwarzian, see \cite{Mertens:2022irh} for a comprehensive review, we compute the one-loop contribution to the non-relativistic Schwarzian theory and find that it correctly reflects the number of generators of EdS-G. On the bulk side we present a Newton-Cartan incarnation that geometrises the EdS-G algebra and we show that its solves the equations of motion of a non-relativistic version of JT. 

The structure of the paper is as follows: We compute one-loop fluctuations to the boundary theory governing the dynamics of a two-dimensional EdS-G gravity theory in Section \ref{sec:boundary}. We continue with a further presentation of a Newton-Cartan formulation of the gravitational bulk in Section \ref{sec:bulk}. In the last Section \ref{sec:discuss}, we relate our computations with recent developments in the fields and discuss future prospects.
\section{The boundary description of non-relativistic de Sitter space}\label{sec:boundary}
We study the Schwarzian-type effective boundary action associated to a non-relativistic bulk theory whose symmetries are governed by the two-dimensional Galilean de Sitter algebra. This action was derived in \cite{Gomis:2020wxp} using a first-order formulation in terms of BF variables, and it was shown that this governs the boundary dynamics of an EdS-G gravity theory. We begin by reviewing such a derivation. Because the Galilean de Sitter algebra is isomorphic to the Carrollian Anti-de Sitter algebra, a similar boundary action also appears in Carrollian settings of \cite{Grumiller:2020elf}, although the bulk interpretation is different. This section is organised in the following manner. We first introduce the action in section \ref{sec:GdS-boundaryaction}, we then analyse the symmetries of the action in section \ref{sec:isometries}, determine its classical saddle in section \ref{sec:eom}, and finally evaluate its one-loop contribution in section \ref{sec:loop}.
\subsection{The Galilean de Sitter boundary action} \label{sec:GdS-boundaryaction}
Let us start by considering the two-dimensional de Sitter space symmetry algebra  $\mathfrak{so}(1,2)$ whose only non-vanishing commutation relations are
\begin{align}\label{eq:dSalgebra}
	[B,\tilde H] =\tilde P\,, \qquad
	[B,\tilde P] = \tilde H\,, \qquad
	[\tilde H,\tilde P] = -\tilde \Lambda \, B\,.
\end{align}
Here $B$ generates Lorentzian boosts, $\tilde P$ spatial translations and $\tilde H$ time translations. Here $\tilde \Lambda$ denotes the cosmological constant. The tilde differentiates with generators that we introduce after contraction. Restricting to a positive value of $\tilde \Lambda$, one possible Inönü--Wigner contraction in which one morally takes the speed of light to infinity while simultaneously rescaling $\tilde{\Lambda}$ leads to the non-relativistic EdS-G algebra, which is sometimes referred to as the Newton-Hooke$_+^2$ algebra \cite{Bacry:1968zf}, given by\footnote{Notice that despite of the absence of the speed of light $c$ in the symmetry algebra \eqref{eq:dSalgebra}, a careful incorporation of $c$ factors modifies the commutator between momenta as $[\tilde H,\tilde P] = c^2\,\tilde \Lambda\, B$. Then, going through the contraction process, it is necessary to take simultaneously $c\to \infty$ and $\tilde \Lambda \to 0$ limits; otherwise, we would have to delete the boost $B$ from the algebra. The origin of $c$ factors arises from an analysis of the equations governing a non-relativistic cosmological model, based on geodesics in de Sitter space \cite{Gibbons:2003rv}. In this paper we use units in which $ c=1$.} 
\begin{align}\label{eq:GaldSalgebra}
	[G,H] = P\,, \qquad
	[G,P] = M\,, \qquad
	[H,P] = -\Lambda\, G\,,
\end{align}
where $G$ denotes the Galilean boost, $H$ an $P$ spacetime translational generator, and $M$ a central generator. The EdS-G algebra is the analogue of the Bargmann algebra in the presence of a universal cosmological repulsion or attraction $\tilde{\Lambda}= 1/\ell^2$. Here $\ell$ stands for the characteristic scale. Given this algebra, we want to formulate a bulk and boundary theory that realises the algebra. We will now focus on engineering the boundary theory.

Within this limit, we propose to formulate the gravity model.
For later purposes, it is convenient to introduce the conformal basis
\begin{align}
	\mathcal D = \ell H\,, \qquad
	\mathcal H = \ell P + G\,, \qquad
	\mathcal K = \ell P - G\,, \qquad
	\mathcal Z = \ell M\,.
\end{align}
 In this basis, the algebra \eqref{eq:GaldSalgebra} becomes
\begin{align} \label{eq:conformal-galilean}
	[\mathcal H,\mathcal D] = \mathcal H\,, \qquad
	[\mathcal K,\mathcal D] = -\mathcal K\,, \qquad
	[\mathcal H,\mathcal K] = 2\mathcal Z\,,
\end{align}
with a non-degenerate invariant bilinear form, given by
\begin{align} \label{eq:pairing}
	\langle \mathcal D,\mathcal D\rangle = c_0\,, \qquad
	\langle \mathcal D,\mathcal Z\rangle = c_1\,, \qquad
	\langle \mathcal H,\mathcal K\rangle = -2c_1\,.
\end{align}
Here $c_0$ and $c_1\neq 0$ are two arbitrary constants. Notice that the algebra \eqref{eq:conformal-galilean} is exactly the extended two-dimensional version of the Poincaré algebra (also known as the Maxwell algebra).
The EdS-G algebra \eqref{eq:GaldSalgebra} was explicitly realised in \cite{Gomis:2020wxp}, where both in BF formalism and metric formalism an EdS-G gravity model was constructed along with the boundary action. In the remainder of this subsection we recap results relevant to the current work. In this context, the construction of the boundary action for a two-dimensional Galilean gravity theory can be interpreted as the action principle for a particle moving on the group manifold. The procedure goes as follows. Consider, the left-invariant Maurer-Cartan form $\Omega$ associated to the \eqref{eq:GaldSalgebra}, which is built from the group element 
\begin{align}
	U = e^{s \mathcal Z} e^{\rho \mathcal H} e^{y \mathcal K} e^{u \mathcal D}\,,
\end{align}
where $\{s, \rho ,y,u\}$ are coordinates of the group manifold. Then, we compute the associated (left-invariant) Maurer--Cartan one-form given by
\begin{align}
	\Omega =U^{-1}dU=
	(ds + 2y\,d\rho)\,\mathcal Z
	+ e^{u}d\rho\,\mathcal H
	+ e^{-u}dy\,\mathcal K
	+ du\,\mathcal D\,.
\end{align}
Using the invariant bilinear form \eqref{eq:pairing}, we construct the action principle as
\begin{align}
	S[s,\rho,y,u]
	 =\int dt\, \Big\langle \Omega, \Omega\Big\rangle^\star= \int dt\Big(
	c_0\,u'^2
	+
	c_1\Big(u's' + 2y\,\rho' - 2\rho' y'\Big)
	\Big)\,.
\end{align}
with prime derivative with respect to the worldline parameter $t$ of the particle, and $\star$ denotes the pull-back action.
This action principle depends on four dynamical fields. Like in $(2+1)$-gravity theory with the reduction from WZNW model to Liouville theory \cite{Coussaert:1995zp}, here it is also possible to further reduce the number  of dynamical fields by imposing the following inverse Higgs constraints (IHC) \cite{Ivanov:1975zq,Gomis:2000bd}
\begin{align}
	\Omega_\mZ = 0\,, \hspace{1cm} \Omega_{\mH} = 1\,,
\end{align}
from which it follows that $y = -\frac{1}{2} e^{u}s'$ and $\rho' =  e^{-u}$. Then, the boundary action becomes
\begin{align}
	S[s,u] = \int dt \Big(  c_0 {u'}^2 + c_1 (s' u' + s'') \Big)\,.
\end{align}
 At the same time, the reduced Maurer--Cartan form or gauge connection becomes
\begin{align}
	a = \Omega\big|_{\text{IHC}}
	= \mathcal H
	- \frac{1}{2}(s'' + s'u')\,\mathcal K
	+ u'\,\mathcal D\,.
\end{align}
As an extra definition, it is convenient to perform the field redefinition $u = v - \log(v')$, so that the reduced action takes the form
\begin{align} \label{eq:Galilean-de-Sitter}
	S[s,v]
	=
	\int dt\left[
	c_0\left(v'^2 - 2v'' + \left(\frac{v''}{v'}\right)^2\right)
	+
	2 c_1\left(s'' + s'\left(v' - \frac{v''}{v'}\right)\right)
	\right]\,.
\end{align}
This one-dimensional action principle is invariant under the EdS-G (global) symmetries \eqref{eq:GaldSalgebra}, and governs the asymptotic dynamics of a two-dimensional EdS-G gravity model. As we will see below, it also admits an invariance under the warped Virasoro algebra. In fact, by considering the one-form gauge connection
\begin{align} \label{eq:assymptotic-gauge-condition}
	a(t)=a_t(t)\, dt \,, \hspace{1cm} a_t(t)
	=
    \mathcal H
	+\mathcal{L}(t)\, \mathcal K
	+
	\mathcal{T}(t)\,\mathcal{D} \,,
\end{align}
the infinite-dimensional symmetry appears from examining the gauge symmetries preserving this structure for $a$. Here the functions $\mathcal{L}$ and $\mathcal{T}$ are two arbitrary functions of $t$ only related to the previous fields as follows
\begin{align} \label{eq:LandT}
	\mathcal L(t)
	&=
	-\frac{1}{2}\left(
	s'' + s'\left(v' - \frac{v''}{v'}\right)
	\right)\,, \hspace{1cm}
	\mathcal T(t)
	=
	v' - \frac{v''}{v'}\,,
\end{align}
These relations arise naturally from studying the boundary dynamics of the BF model in a two-dimensional bulk. Concretely, the relations are supported through the equation $a=U^{-1} dU$. We find that the gauge symmetry transformation laws that preserve the structure of $a$, demand that
\begin{align} \label{eq:transformation-laws1}
	\delta \mathcal L
	&=
	\sigma \mathcal L' + 2\mathcal L \sigma' + \mathcal T\chi' + \chi''\,,
	\\
	\delta \mathcal T
	&=
	\sigma' \mathcal T + \sigma \mathcal T' - \sigma''\,,
\end{align}
with $\chi$ and $\sigma$ two independent gauge parameters which are functions of $t$. 
These are precisely the transformation laws underlying the warped Virasoro structure of the theory \cite{Afshar:2019axx}.
Furthermore, the algebra of the canonical generators can be obtained from the identity $\{Q_{\epsilon_1},Q_{\epsilon_2}\}=\delta_{\epsilon_2}Q_{\epsilon_1}$ together with the transformation laws in \eqref{eq:transformation-laws1}. In this case, we have two canonical generators associated with the two gauge  symmetry parameters and their equal-time Poisson brackets are given by
\bseq \label{eq:warped-virasoro}
\ba
i\{ \mathcal{L}_m \,, \mathcal{L}_n \} & =& (m-n)\mathcal{L}_{m+n}+c\,n^3\delta_{n+m, 0}\,,\\
i\{ \mathcal{L}_m \,, \mathcal{T}_n \} & =&-n\mathcal{T}_{m+n} + i \kappa \, m^2 \delta_{n+m, 0}\,,\\
i\{ \mathcal{T}_m \,, \mathcal{T}_n \} & =& 0\,,
\ea
\eseq
with the central charges $c=2\pi c_1$ and $\kappa = 4\pi c_0$. After the redefinitions
$\mathcal{T}_m\to \mathcal{T}_m - i\kappa \delta_m$ and
$\mathcal{L}_m\to \mathcal{L}_m + \frac{c}{2}\delta_m$, we recognise this as the standard warped Virasoro algebra (see \cite{Afshar:2019tvp} for an overview). Therefore, we have shown that the asymptotic structure of the EdS-G gravity model is governed by the warped Virasoro algebra with two different central charges. The global part of the symmetry algebra \eqref{eq:warped-virasoro} corresponding to \eqref{eq:GaldSalgebra} is built on four generators $\{L_0, L_1, P_1, P_{-1} \}$ that preserve the classical vacuum configuration 
\begin{equation}\label{eq:vacuum}
\mathcal{L} = 0 \,, \hspace{0.5cm} \text{and}  \hspace{0.5cm} \mathcal{T} = \text{constant} \,.
\end{equation}
Having made contact with the warped Schwarzian theory in \cite{Afshar:2019tvp}, let us briefly explain in which respects it differs from the present Galilean de Sitter boundary theory. On the one hand, the warped Schwarzian theory is formulated in terms of coadjoint orbits of the warped Virasoro group and the associated Kirillov--Kostant--Souriau symplectic form \cite{Kirillov:1976uta,Kostant:1970qur,Souriau:1970sdd}. The relation should, however, be understood at the level of asymptotic symmetry structure rather than as an identification of the two boundary actions. In reference \cite{Afshar:2019tvp}, the theory is constructed directly from the coadjoint orbit, whereas here the boundary action arises from a non-relativistic de Sitter bulk construction after imposing inverse Higgs constraints. Thus, although the asymptotic symmetry algebra is the same, the bulk origin of the two theories is different. We will also derive the symplectic form entering the path integral measure directly from the effective action by introducing Ostrogradsky canonical momenta, see e.g. \cite{Woodard:2015zca}. To the best of our knowledge, this action-based derivation of the symplectic form has not been emphasised in the literature. As a consistency check, Appendix~\ref{sec:Ostro-sch} shows that the same method reproduces the familiar reduced symplectic structure of the Schwarzian theory. We then revisit the computation of \cite{Afshar:2019tvp} in Appendix~\ref{sec:Ostro-warped-sch} using our formalism. This reproduces the corresponding one-loop result and clarifies in what sense the present Galilean de Sitter boundary action is closely related to, but not identical to, the warped Schwarzian theory.

\subsection{Isometries of the Galilean de Sitter action} \label{sec:isometries}

In order to understand the structure of the path integral and, in particular, the origin of zero modes in the one-loop analysis, we first determine the global symmetries of the boundary action. Having reviewed the derivation of the Galilean de Sitter action in equation \eqref{eq:Galilean-de-Sitter}, we recall that it is given by
\begin{align} \label{eq:EdS-G}
    I_{\mathrm{EdS-G}} = \int dt \Bigg[
        c_0 \left(
            v'^2 - 2v'' + \left(\frac{v''}{v'}\right)^2
        \right)
        +
        2 c_1 \left(
            s'' + s' \left( v' - \frac{v''}{v'} \right)
        \right)
    \Bigg]\,.
\end{align}
We now observe that the action is invariant under the finite transformations
\begin{align}
    s(t) &\mapsto s(t) + a + b e^{-v(t)}\,, \hspace{1cm}
    v(t) \mapsto -\log\!\big(c + d e^{-v(t)}\big)\,,
\end{align}
where $a,b,c,d$ are arbitrary constants. The corresponding infinitesimal isometries which leave the action invariant are given by
\begin{align} \label{eq:infinitesimal-isometries}
    \delta s &= \epsilon_a + \epsilon_b e^{-v(t)}\,, \hspace{1cm}
    \delta v = \epsilon_c + \epsilon_d \, e^{v(t)}\,,
\end{align}
where $\{\epsilon_a, \epsilon_b, \epsilon_c, \epsilon_d\}$ are infinitesimal parameters, which will be used to construct the corresponding Noether charges. These four independent transformations generate the global part of the warped Virasoro symmetry algebra \eqref{eq:warped-virasoro} and will later be associated with the non-Gaussian sector of the path integral. Because the action contains higher-derivative terms, the corresponding Noether charges must be constructed using the generalised Ostrogradsky formalism. We compute the generalised Ostrogradsky momenta as
\begin{align}
    p^{(s)}_{1}
    &=
    \frac{\partial L}{\partial s'}
    -
    \frac{d}{dt}\!\left(\frac{\partial L}{\partial s''}\right)
    =
    2c_1 \left(v'
    -
    \frac{ v''}{v'} \right)\,,
    \\
    p^{(v)}_{1}
    &=
    \frac{\partial L}{\partial v'}
    -
    \frac{d}{dt}\!\left(\frac{\partial L}{\partial v''}\right)
    =
    2c_1 s'
    +
    2c_0 v'
    + \frac{2}{v'}\left[ c_1 s'' + c_0\left(\frac{{v''}^2}{{v'}^2} - \frac{v'''}{v'} \right) \right]\,,
    \\
    p^{(s)}_{2}
    &=
    \frac{\partial L}{\partial s''}
    =
    2c_1\,,
    \\
    p^{(v)}_{2}
    &= 
    \frac{\partial L}{\partial v''}
    = -\frac{2}{v'}\left[c_1 s' +  c_0 \left(v' - \frac{v''}{v'}\right) \right]\,.
\end{align}
From these, we may compute the four associated charges, obtaining
\begin{align}
    Q[\epsilon]
    =
   \int dt\Big[ p^{(s)}_{1}\,\delta s
    +
    p^{(v)}_{1}\,\delta v
    +
    p^{(s)}_{2}\,\delta s'
    +
    p^{(v)}_{2}\,\delta v' \Big]\,,
\end{align}
where $\delta s$ and $\delta v$ are the infinitesimal symmetry variations given in \eqref{eq:infinitesimal-isometries}. Substituting these variations into the charge formula yields
\begin{align}
    Q_a &= Q[\epsilon_a] = 2c_1 \int dt \left(v' - \frac{v''}{v'}\right) \,,  \\
    Q_b &= Q[\epsilon_b] = -2c_1 \int dt\frac{v''}{v'}e^{-v(t)}  \,, \\
    Q_c &= Q[\epsilon_c] = 2 \int \frac{dt}{v'}\left[ c_1 s'' + c_0 \left( \frac{{v''}^2}{{v'}^2} - \frac{v'''}{v'} \right) \right]\,, \\
    Q_d &= Q[\epsilon_d] = 2\int \frac{dt \, e^{v(t)}}{v'}\left[ c_1 s'' +  c_0 \left(v'' +  \frac{{v''}^2}{{v'}^2} - \frac{v'''}{v'}  \right) \right]\,.
\end{align}
Evaluating the Poisson brackets of the charges directly, we find that $Q_a,Q_b,Q_c,Q_d$ realise the conformal Galilean algebra in \eqref{eq:conformal-galilean}. More explicitly, the non-vanishing brackets are
\begin{align}
    \{Q_c,Q_b\}=-Q_b\,, \hspace{1cm} \{Q_c,Q_d\}=Q_d \,, \hspace{1cm} \{Q_d,Q_b\}=Q_a\,.
\end{align}
Identifying $Q_c = \mathcal{D}$, $Q_b = \mathcal{H}$, $Q_d = \mathcal{K}$, and $Q_a = 2\mathcal{Z}$, we recover the algebra \eqref{eq:conformal-galilean}. This confirms that the conserved charges of the boundary theory realise the conformal Galilean algebra in its global symmetries. As we will see in the next sections, these symmetries are directly responsible for the presence of zero modes in the quadratic fluctuation analysis, which in turn require a separate treatment in the path integral.

\subsection{Equations of motion and saddle point}\label{sec:eom}

We now determine the classical saddle of the action and analyse quadratic fluctuations around it, which will form the basis of the one-loop computation in the next section. To this end, we decompose the fields into background configurations and fluctuations
\begin{equation}\label{eq:fluctuations}
	v
	=
	\overline{v}
	+
	\delta v
	\,,
	\hspace{1cm}
	s
	=
	\overline{s}
	+
	\delta s
	\,,
\end{equation}
where the barred quantities denote the background fields and the quantities accompanied by $\delta$ denote the fluctuations. 
The background fields are taken to satisfy the boundary conditions
\begin{align} \label{eq:periodicity}
    \bar{v}(t + \beta) = v(t) - 2\pi i \,, \hspace{1cm} \bar{s}(t+\beta) = \bar{s}(t)\,, 
\end{align}
while the fluctuations are $\beta$-periodic. Notice that from now on, we have implicitly promoted the coordinate $t$ to be cyclic. The appearance of the imaginary shift in $\bar{v}$ might not be surprising, at least on an algebraic level, when we consider the very similar actions presented in \cite{Grumiller:2020elf,Afshar:2019tvp}, which are invariant under the Extended AdS-Carroll algebra. This algebra is isomorphic to EdS-G by interchanging the spacetime translational generators as $P \leftrightarrow H$. The actions there, in essence, have $v\to i v$ as compared to the EdS-G action. In the next section, we will require the fluctuations around the saddle-point solution of $v$ to be real; as such, $v'$ will turn out to receive only a complex shift due to the boundary conditions above. We believe that this is reminiscent of the choices of complex contours in discussions surrounding the treatment of the boundary theory of two-dimensional \emph{relativistic} de Sitter \cite{Maldacena:2019cbz,Cotler:2019nbi}.

Having specified the boundary conditions, we now expand the action to first order in the fluctuations. Requiring the first variation to vanish, then yields the equations of motion for the background fields. We obtain
\begin{align}
    I^{(1)}_{\mathrm{EdS-G}} &= -2\int dt \left\{
        \partial_t \left[
            c_1\left(\bar{s}' + \frac{\bar{s}''}{\bar{v}'} \right)
            +
            c_0 \left(
                \bar{v}'
                - \frac{\bar{v}'''}{\bar{v}'^2}
                - \frac{\bar{v}''^2}{\bar{v}'^3}
            \right)
        \right] \delta v
        +
        c_1 \partial_t \left(
            \bar{v}' - \frac{\bar{v}''}{\bar{v}'}
        \right) \delta s
    \right\} \,.
\end{align}
From this, we may read off the equations of motion, which are given by
\begin{align}
  \delta v:& \hspace{1cm}  \partial_t \left[             c_1\left(\bar{s}' + \frac{\bar{s}''}{\bar{v}'} \right)
            +
            c_0 \left(
                \bar{v}'
                - \frac{\bar{v}'''}{\bar{v}'^2}
                - \frac{\bar{v}''^2}{\bar{v}'^3}
            \right) \right] = 0\,, \\
  \delta s:& \hspace{1cm} \partial_t\left(  \bar{v}' - \frac{\bar{v}''}{\bar{v}'}  \right) = 0\,.
\end{align}
The equations can be solved sequentially: first, one determines $\bar v$ from the second equation, and then substitutes this result into the first equation to determine $\bar s$. The solutions are then given by
\begin{equation}\begin{aligned}
    e^{-\overline{v}}
    =&~
    k_{1} e^{k_{2} t}+k_{3}
    \,,
    \\
    \overline{s}'
    =&~
    k_{4} e^{\overline{v}/c_{1}}
    +
    k_{5}
    \,,
\end{aligned}\end{equation}
with $k_{1},...,k_{5}$ integration constants. Imposing the periodicity constraints in \eqref{eq:periodicity} we choose the saddle point to be
\begin{align} \label{eq:saddle-sol}
  \bar{s}(t) = 0\,, \hspace{1cm}  \bar{v}(t) = -\frac{2 \pi i}{\beta} t\,,
\end{align}
in agreement with the values appearing in \eqref{eq:vacuum}, which are invariant under the global conformal Galilean subalgebra (see Appendix \ref{ap:global-symmetries-twised}).
Constant shifts in $\bar s$ and $\bar v$ have been omitted, since the action depends only on derivatives of these fields.
Continuing with higher orders in the expansion, the action to second order in the fluctuations gives
\begin{align} \label{eq:second-order-fluc-GdS-action}
    I^{(2)}_{\mathrm{EdS-G}}
    &= \int dt \Bigg[
        2 c_1 \,\delta s' \,\delta v'
        + c_0 {\delta v'}^2
        + \frac{2 c_1}{\bar{v}'} \,\delta v' \,\delta s''
        + \frac{c_0}{\bar{v}'^2} \delta v''^2 
    \nonumber \\[2pt]
    &\hspace{1.5cm}
        - \frac{2 c_1 \bar{s}' \bar{v}''}{\bar{v}'^3} \delta v'^2 + \frac{3 c_0 \bar{v}''^{2}}{\bar{v}'^4} \delta v'^2
        + \frac{2 c_1 \bar{s}'}{\bar{v}'^2} \,\delta v' \,\delta v''
        - \frac{4 c_0 \bar{v}''}{\bar{v}'^3} \,\delta v' \,\delta v''
    \Bigg] \,.
\end{align}
To fully account for the quadratic fluctuations, we must evaluate the path integral over the fluctuation modes, which we will do in the next section. In addition, the path integral measure is fixed as the Pfaffian of the associated symplectic form, defined as
\begin{equation}\label{eq:sympfo}
	\Omega 
	=
	\int^{\beta}_{0}dt
	\Big(
	dp_{v}\wedge dv
	+
	dp_{s}\wedge ds
	\Big)
	\,.
\end{equation}
There the generalized momenta $p_{v}$ and $p_{s}$ are defined, using the Ostrogradsky formalism, through
\begin{equation}
	p_{w}
	=
	\frac{\partial \mathcal{L}}{\partial w'}
	-
	\frac{d}{dt}
	\left(
	\frac{\partial \mathcal{L}}{\partial w''}
	\right)
	\,,
\end{equation}
with $\mathcal{L} = \mathcal{L}(w, w', w'')$ the Lagrangian corresponding to the action in \eqref{eq:EdS-G}, which depends at most on second derivatives, and where $w$ denotes either $v$ or $s$. Applying this to the quadratic fluctuation action, we obtain the corresponding generalised Ostrogradsky momenta
\begin{align}
    p_{\delta s} &=  \frac{2c_1}{\bar{v}'}\left( \left(\bar{v}' - \tfrac{\bar{v}''}{\bar{v}'}\right)\delta v' - \delta v'' \right) \,, \\
    p_{\delta v}  
    &=\frac{2c_1}{\bar{v}'}\left( \bar{v}' \delta s' + \delta s'' - \tfrac{\bar{s}''}{\bar{v}'}\delta v' \right)
    + \frac{2c_0}{\bar{v}'^2}\left( \left(\bar{v}'^2 - \tfrac{3 \bar{v}''^2}{\bar{v}'^2} + \tfrac{2 \bar{v}'''}{\bar{v}'} \right) \delta v' + \tfrac{2 \bar{v}''}{\bar{v}'} \delta v'' - \delta v''' \right)\,.
\end{align}
Using these expressions, one finds that the symplectic form is given by
\begin{align} \label{eq:second-order-fluc-GdS-symplectic}
    \Omega_{\mathrm{EdS-G}}^{(2)} &= 2c_1 \int \frac{dt}{\bar{v}'} \left[2 \bar{v}' d\delta v' \wedge d\delta s + d\delta v' \wedge d\delta s' + d\delta s'' \wedge d\delta v - \tfrac{\bar{s}''}{\bar{v}'} d\delta v' \wedge d\delta v \right]  \nonumber  \\ 
    & \hspace{1cm}+ 2c_0 \int \frac{dt}{\bar{v}'^2} \left[  \bar{v}'^2 d\delta v' \wedge d\delta v - d\delta v''' \wedge d\delta v + d\delta v'' \wedge d\delta v'\right]\,.
\end{align}
We are primarily interested in evaluating these expressions on the saddle \eqref{eq:saddle-sol}, namely the quadratic action \eqref{eq:second-order-fluc-GdS-action} and the symplectic form \eqref{eq:second-order-fluc-GdS-symplectic}. Substituting the saddle \eqref{eq:saddle-sol} into these expressions, we obtain
\begin{align} 
    I^{(0)}_{\mathrm{EdS-G}} &= -\frac{4 \pi^2 c_0}{\beta} \,, \label{eq:on-shell-values}\\ 
    I^{(2)}_{\mathrm{EdS-G}} &= 2c_1\int dt \left[  \delta s' \delta v' + i\tfrac{\beta}{2\pi} \delta v' \delta s''  \right]+ c_0\int dt \left[ {\delta v'}^2 - \left( \tfrac{\beta}{2\pi}\right)^2 {\delta v''}^2 \right]\,,\label{eq:secondorderIdS-Gaction} \\
 \nonumber    \Omega^{(2)}_{\mathrm{EdS-G}} &= 4c_1\int dt \left[ d\delta v' \wedge d\delta s + i \tfrac{\beta}{2\pi} d\delta s'' \wedge d\delta v  \right]  \\ &\hspace{1.5cm} 2c_0\int dt\left[ d\delta v' \wedge d\delta v + \left( \tfrac{\beta}{2\pi} \right)^2 d\delta v''' \wedge d\delta v \right]\,,\label{eq:secondorderIdS-Gomega} 
\end{align}
where $I^{(1)}_{\mathrm{EdS-G}} = 0$ by the equations of motion.

\subsection{Evaluating the one-loop path integral}\label{sec:loop}
In this section, we evaluate the one-loop contribution to the boundary path integral for the Galilean de Sitter gravity theory. We will do this in the boundary description.
\emph{Exactly} equating a gravitational path integral to the path integral of a quantum mechanical system in one dimension -- its boundary description -- is highly non-trivial, but well-studied in JT gravity \cite{Stanford:2017thb}.
This can be understood in the following way: since the dilaton in JT gravity acts as a Lagrange multiplier, one can perform integration over the dilaton at the level of the path integral, which fixes the curvature of the geometry and only leaves boundary fluctuations. We will return later to the bulk formulation of non-relativistic JT gravity in section~\ref{sec:bulk}, where we will argue that an analogous mechanism is at work for Galilean de Sitter gravity, which can be cast into a JT-like action formulation with a dilaton that again acts as a Lagrange multiplier and may be integrated out. The relevant path integral we want to solve becomes
\begin{equation}
	\mathcal{Z}_{\text{EdS-G}}
	=
	\int_{\mathcal{M}}
	[\mathcal{D}s]
	\,
	[\mathcal{D}v]
	\,
	e^{-I_{\text{EdS-G}}[s,v]}
	\,.
\end{equation}
Given the boundary equations we impose, the space $\mathcal{M}$ we integrate over consists of all configurations of $s$ and $v$ subject to the boundary conditions \eqref{eq:periodicity}, and with EdS-G symmetry modded out. Plugging in the fluctuations, this yields
\begin{equation}
	\mathcal{Z}_{\text{EdS-G}}
	=	
	e^{-I^{(0)}_{\text{EdS-G}}[\overline{s},\overline{v}]}
	\int_{\mathcal{M}}
	[\mathcal{D}\delta s]
	\,
	[\mathcal{D}\delta v]
	\,
	e^{-I^{(2)}_{\text{EdS-G}}[\delta s,\delta v]}
	\,.
\end{equation}
As we have seen, the theory \eqref{eq:EdS-G} possesses a set of global symmetries generated by the EdS-G algebra \eqref{eq:GaldSalgebra}. 
These will play a crucial role in evaluating the path integral, as they give rise to zero modes.
The evaluation of the path integral proceeds in three steps.
First, we expand the fluctuations in Fourier modes, which diagonalise the quadratic action.
Second, we evaluate the resulting Gaussian integrals for the non-zero modes.
Third, we treat the zero modes separately, as they are associated with the global symmetries identified in Section \ref{sec:isometries}.

To implement the first step, we expand the fluctuations in Fourier modes as follows
\begin{equation}\label{eq:fourier}
	\delta v(t)
	=
	\sum_{n \in \mathbb Z } \Big( v_{n}^{(R)}+ iv_{n}^{(I)} \Big)\,  e^{\frac{2\pi}{\beta}in t}
	\,,
	\hspace{1cm}
	\delta s(t)
	=
	\sum_{n \in \mathbb Z } \Big( s_{n}^{(R)}+ is_{n}^{(I)} \Big)\,e^{\frac{2\pi}{\beta}in t}	
	\,,
\end{equation} 
where we will require the real ($R$) and imaginary parts ($I$) to satisfy
\begin{equation}
	w_{n}=w_{n}^{(R)}+i w_{n}^{(I)}
	\,,
	\quad
	w_{-n}^{(R)}=w_{n}^{(R)}
	\,,
	\quad
	w_{-n}^{(I)}=-w_{n}^{(I)}
	\,,
\end{equation}
and $w$ is a placeholder for $s$ and $v$. Plugging the Fourier expansion into the second-order on-shell action \eqref{eq:secondorderIdS-Gaction} and symplectic form \eqref{eq:secondorderIdS-Gomega} yields
\begin{align}
    I^{(2)}_{\mathrm{EdS-G}} &= -\frac{4\pi^2}{\beta} \sum_{n\in \mathbb Z} \Big( 2c_1 n^2(n-1) \delta s_n \delta v_{-n} + c_0 n^2(n^2-1) \delta v_n \delta v_{-n} \Big)\,, \\
    \Omega^{(2)}_{\mathrm{EdS-G}} &= 4 \pi i \sum_{n\in \mathbb Z} \Big( 2 c_1 n (n-1) d\delta v_n \wedge d\delta s_{-n} - c_0 n (n^2-1) d\delta v_n \wedge d\delta v_{-n}\Big)\,.
\end{align}
We have used $\int^{\beta}_{0}dt e^{i(m-n)t}=\beta \, \delta_{mn}$.
Using the real and imaginary parts of the Fourier coefficients, we may rewrite the expressions entirely in terms of positive modes, in which case we find
\begin{align}
\nonumber      I^{(2)}_{\mathrm{EdS-G}} &= \frac{16 \pi^2}{\beta}c_1 \delta s_{-1} \delta v_{1} -\frac{8 \pi^2}{\beta}c_0 \sum_{n \geq 2} n^2(n^2-1)\left[ \delta v_n^{(R)} \delta v_n^{(R)} + \delta v_n^{(I)} \delta v_n^{(I)} \right] \\
 \nonumber    &\hspace{1cm} -\frac{16 \pi^2}{\beta} c_1 \sum_{n \geq 2} n^2\left[ \delta s_n^{(R)}\delta v_n^{(R)} + \delta s_n^{(I)} \delta v_n^{(I)} \right] \\
    &\hspace{1.5cm}- \frac{16 \pi^2 i}{\beta} c_1 \sum_{n \geq 2} n^3 \left[ \delta s_n^{(R)}\delta v_n^{(I)} - \delta s_n^{(I)} \delta v_n^{(R)} \right]\,.
\end{align}
For the symplectic form, we get that
\begin{align}
\nonumber     \Omega_{\mathrm{EdS-G}}^{(2)} &= 16 \pi i c_1 d\delta v_{-1} \wedge d\delta s_{1} + 16 \pi i c_1 \sum_{n \geq 2} n^2\left[ d\delta v_n^{(R)} \wedge d\delta s_n^{(R)} + d\delta v_n^{(I)} \wedge d\delta s_n^{(I)} \right]  \\
\nonumber     &\hspace{1cm} - 16 \pi c_1 \sum_{n \geq 2} n \left[ d\delta v_n^{(I)} \wedge d\delta s_n^{(R)} - d\delta v_n^{(R)} \wedge d\delta s_n^{(I)} \right] \\ &\hspace{1.5cm} + 32 \pi  c_0 \sum_{n \geq 2} n (n^2-1) d\delta v_n^{(I)} \wedge d\delta v_n^{(R)}\,.
\end{align}
We have explicitly separated the $n=1$ mode, as it requires special treatment and is associated with the global symmetries identified in Section~\ref{sec:isometries}. For $n \geq 2$, the quadratic fluctuation integral is Gaussian and can therefore be evaluated using 
\begin{align}
    \int_{\mathbb{R}^4} e^{-\frac{1}{2}x^T A_n x} = \sqrt{\frac{(2\pi)^4}{\det A_n}}\,.
\end{align}
For each mode $n \geq 2$, the quadratic action may be written in the form $\frac{1}{2}x^{T}A_{n}x$, where the fluctuation vector is
$x=\big(\delta v_{n}^{(R)},\delta v_{n}^{(I)},\delta s_{n}^{(R)},\delta s_{n}^{(I)}\big)$. For $n \geq 2$, the matrix $A_n$ takes the form
\begin{align}
A_n
&= \frac{16\pi^2 n^2 }{\beta}
\begin{pmatrix}
c_0 (n^2-1) & 0 & 2c_1 & 2i c_1 n \\
0 & c_0 (n^2-1) &  -2i c_1 n & 2 c_1 \\
2c_1 & 2i c_1 n & 0 & 0 \\
-2i c_1 n & 2 c_1 & 0 & 0
\end{pmatrix}\,.
\end{align}
The corresponding symplectic form may likewise for $n \geq 2$ be represented by an $4\times 4$ antisymmetric matrix $\Omega_{n}$
\begin{align}
    \Omega_n = 16 \pi i \begin{pmatrix}
        0 & -2 i c_0 n (n^2-1) & c_1 n^2 & -c_1 i n \\
        2 i c_0 n (n^2-1) & 0 & c_1 i n & c_1 n^2 \\
        -c_1n^2 & -c_1 in  & 0 & 0 \\
        c_1 i n & -c_1 n^2 & 0  & 0 
    \end{pmatrix}\,.
\end{align}
From these expressions, we obtain
\begin{align}
 \sqrt{\frac{(2\pi)^4}{\det A_n}} = \frac{\beta^2}{256 \pi^2 c_1^2 n^4 (n^2-1)}\,, \hspace{1cm} \mathrm{Pf}(\Omega_n) = 256 \pi^2 c_1^2 n^2(n^2-1)\,.
\end{align}
Hence we find
\begin{align}\label{eq:Zneedsregulator}
 Z(\beta) =   e^{\frac{4\pi^2 c_0}{\beta}}\mathcal{B}(\beta) \prod_{n \geq 2} \mathrm{Pf}(\Omega_n) \sqrt{\frac{(2\pi)^4}{\det A_n}} =  e^{\frac{4\pi^2 c_0}{\beta}}\mathcal{B}(\beta) \prod_{n \geq 2} \frac{\beta^2}{n^2} \,.
\end{align}
Before detailing $\mathcal{B}(\beta)$, we evaluate the infinite product using zeta-function regularisation. 
This relies on the analytic continuation of the Riemann zeta function,
\begin{align}
\zeta(s)=\sum_{n\ge1} n^{-s},
\qquad
\zeta'(s)=-\sum_{n\ge1} n^{-s}\log n\,.
\end{align}
Applying this to the product in \eqref{eq:Zneedsregulator}, we obtain
\begin{align}
\prod_{n\ge2}\frac{\beta^2}{n^2}
&=
\exp \Bigg(\sum_{n\ge2}\bigl(2\log\beta-2\log n\bigr)\Bigg)
=
\exp\Big(2\log(\beta)\,(\zeta(0)-1)+2\zeta'(0)\Big)  = \frac{1}{2\pi\beta^3}\,,
\end{align}
where we used the regularized values $\zeta(0)=-\frac{1}{2}$ and $\zeta'(0)=-\frac{1}{2}\log(2\pi)$. The non-Gaussian contribution from the $n=1$ modes is captured by $\mathcal{B}(\beta)$, which we compute as follows
\begin{align}
\mathcal{B}(\beta)
&=
16\pi i\, c_1
\Bigg(
\iint d\delta v_1^{(R)} d\delta s_1^{(R)}
+
\iint d\delta v_1^{(I)} d\delta s_1^{(I)}
+
i \iint d\delta v_1^{(R)} d\delta s_1^{(I)}
-
i \iint d\delta v_1^{(I)} d\delta s_1^{(R)}
\Bigg)
\nonumber\\
&\hspace{-0.3cm} \times
\exp\Bigg(
\frac{16\pi^2}{\beta} c_1
\Big(
\delta s_1^{(R)} \delta v_1^{(R)}
+
\delta s_1^{(I)} \delta v_1^{(I)}
-
i \left( \delta s_1^{(I)} \delta v_1^{(R)} - \delta s_1^{(R)} \delta v_1^{(I)} \right)
\Big)
\Bigg) = 4 \beta \,,
\end{align}
where in order to evaluate the preceding integral we introduced a regulator $-i \epsilon \abs{v_1}^2$, with $\mathrm{Im}(\epsilon) < 0$ 
to ensure convergence. The integral is first computed at 
finite $\epsilon$, after which the $\epsilon \to 0$ limit is taken, 
noting that the final result is regulator-independent.

Collecting all contributions, we obtain the final expression for the partition function
\begin{equation}
	\mathcal{Z}_{\text{EdS-G}}(\beta)
	=	
	\frac{2}{\pi \beta^{2}}
	\exp
	\Bigg(
		\frac{4\pi^2 c_{0}}{\beta}
	\Bigg)
	\,.
\end{equation}
This corresponds to the main result of this section. The overall scaling $\mathcal{Z}_{\text{EdS-G}}(\beta) \sim \beta^{-2}$ is consistent with the general expectation that each global symmetry generator contributes a factor of $\beta^{-1/2}$. Since the EdS-G algebra possesses four generators, this results in the observed $\beta^{-2}$ behaviour. This is a characteristic feature of Schwarzian-type theories, in which exponential growth arises from the classical saddle, and the quantum fluctuations provide a temperature-dependent prefactor determined by the number of global symmetry generators.

Performing an inverse Laplace transform, we obtain the density of states (dos) as a function of the energy $E$,
\begin{equation}
\mathcal{D}_{\text{dos}}(E)
=
\frac{\sqrt{E}}{\pi^2 \sqrt{c_0}}
\, I_1\!\left(4\pi \sqrt{ c_0 E}\right)\,,
\end{equation}
where $I_1$ is the modified Bessel function of the first kind. In particular, for small energies $E \to 0$, this behaves as
\begin{equation}
\mathcal{D}_{\text{dos}}(E) = \frac{2}{\pi} E + \mathcal{O}(E^2)\,,
\end{equation}
which shows that the density of states vanishes linearly at low energies. The linear behaviour at low energies also reflects how quantum fluctuations resolve the naive semiclassical gap in the spectrum and indicate the presence of low-energy states in the quantum theory.

%
\section{A bulk Newton-Cartan description of Galilean de Sitter}\label{sec:bulk}
%
In the previous section, we developed a boundary description of non-relativistic de Sitter space and evaluated its one-loop contribution. The goal of this section is to complement that analysis by constructing the corresponding bulk description and elucidating its geometric structure. 
A particular novelty is that we work out a non-relativistic equivalent of JT gravity and show that the geometry corresponding to EdS-G satisfies its equations of motion.
We begin in section \ref{subsec:bulk} by deriving the two-dimensional Newton--Cartan geometry associated with the bulk gauge connection. In section \ref{sec:NC} we show that this geometry realises the EdS-G algebra \eqref{eq:GaldSalgebra} as its isometry algebra, and in section \ref{sec:uplifting} we provide a relativistic uplift to three dimensions. We then turn in section \ref{sec:NR-JT} to a complementary second-order description, showing that the same Newton--Cartan geometry arises from a non-relativistic JT-type action. Finally, in section \ref{sec:firstorder} we compare this second-order formulation with the underlying first-order gauge-theoretic description. Taken together, these results establish a bulk description whose geometric and symmetry structures precisely match those of the boundary dynamics studied in section \ref{sec:boundary}.

\subsection{Bulk Newton--Cartan geometry from the gauge connection} \label{subsec:bulk}

To construct a bulk description of Galilean de Sitter gravity, we work in a two-dimensional Newton--Cartan framework. Newton--Cartan geometry, originally introduced by \'Elie Cartan \cite{Cartan:1923,Cartan:1924}, provides a geometric formulation of non-relativistic gravity analogous to the role played by Riemannian geometry in general relativity. In particular, it is the natural geometric language for theories with Galilean symmetry. We refer to \cite{Hartong:2022lsy} for an excellent modern review of non-relativistic gravity, including its history, formulation, and recent developments.

In two dimensions, the Newton--Cartan data consist of a temporal one-form $\tau_\mu$, a spatial vielbein $e_\mu$ together with its inverse $e^\mu$, and a $U(1)$ mass gauge field $m_\mu$. More generally, Newton--Cartan geometry may also be torsionful, depending on whether the clock form $\tau_\mu$ is closed. Such torsionful geometries play an important role in more general non-relativistic settings and can account for effects such as non-relativistic gravitational lensing \cite{Hansen:2019vqf}.

Our starting point is the boundary gauge field derived in Section \ref{sec:boundary}. In the gauge \eqref{eq:assymptotic-gauge-condition}, it takes the form
\begin{align}
    a =  \mathcal{H} + \mathcal{L}(t)\mathcal{K} + \mathcal{T}(t)\mathcal{D} \,.
\end{align}
To reinterpret this in Newton--Cartan variables, we pass from the basis $\{ \mathcal{H}, \mathcal{D}, \mathcal{K}, \mathcal{Z} \}$ to the basis adapted to the EdS-G algebra,
\begin{align}
    H = \frac{1}{\ell}\mathcal{D} \,, \qquad
    P = \frac{1}{2\ell}\bigl(\mathcal{H} + \mathcal{K}\bigr) \,, \qquad
    G = \frac{1}{2}\bigl(\mathcal{H} - \mathcal{K}\bigr) \,, \qquad
    M = \frac{1}{\ell}\mathcal{Z} \,.
\end{align}
We then introduce the radial group element
\begin{align}
    b(r) = e^{-r \mathcal{H}} \,,
\end{align}
and construct the bulk connection in radial gauge according to the permissible gauge transformation
\begin{align}
    A(t,r) = b^{-1}(r) d b(r) + b^{-1}(r) a(t) b(r) \,.
\end{align}
Besides that, expanding the one-form $A$ in the extended Galilean basis \eqref{eq:EdS-G} as
\begin{align}
    A(t,r) = \tau H + e P + \omega G + m M \,,
\end{align}
allows us to read off the Newton--Cartan fields directly from the coefficients of the generators, obtaining
\begin{align}
    \tau_\mu dx^\mu &= \ell \,\mathcal{T}(t)\, dt \,, \\
    e_\mu dx^\mu &= \ell \Bigl(1 + \mathcal{L}(t) + r \mathcal{T}(t)\Bigr) dt - \ell\, dr \,, \\
    m_\mu dx^\mu &= 2 \ell r \mathcal{L}(t)\, dt \,, \\
    \omega_\mu dx^\mu &= \Bigl(1 - \mathcal{L}(t) + r \mathcal{T}(t)\Bigr) dt - dr \,.
\end{align}
These gauge fields define the two-dimensional Newton--Cartan geometry associated with the bulk gauge connection $A(t,r)$. 
It is convenient to introduce the inverse temporal vector $v^\mu$ and the inverse spatial vielbein $e^\mu$, which together with $\tau_\mu$ and $e_\mu$ satisfy the following orthogonality and completeness relations of Newton--Cartan geometry
\begin{align} \label{eq:completenesseq}
    v^\mu \tau_\mu &= 1 \,, &
    \tau_\mu e^\mu &= 0 \,, &
    e_\mu v^\mu &= 0 \,, &
    e_\mu e^\mu &= 1 \,, &
    \delta^\mu{}_\nu &= v^\mu \tau_\nu + e^\mu e_\nu \,,
\end{align}
which we can solve for the contravariant objects, yielding
\begin{align}
    v^\mu \partial_\mu 
    &= \frac{1}{\ell \mathcal{T}(t)} 
    \Bigl( \partial_t + \bigl(1+\mathcal{L}(t)+r\mathcal{T}(t)\bigr)\partial_r \Bigr) \,, \hspace{0.5cm}
    e^\mu \partial_\mu 
    = -\frac{1}{\ell}\,\partial_r \,.
\end{align}
They allow us to construct the degenerate spatial metric and its inverse,
\begin{align}
    h_{\mu\nu} = e_\mu e_\nu \,,
    \qquad
    h^{\mu\nu} = e^\mu e^\nu \,.
\end{align}
For the general connection above, the spatial metric becomes
\begin{align}
    h_{\mu\nu} dx^\mu dx^\nu
    =
    \ell^2 \Bigl( 1+\mathcal{L}(t)+r\mathcal{T}(t) \Bigr)^2 dt^2
    -2\ell^2 \Bigl( 1+\mathcal{L}(t)+r\mathcal{T}(t) \Bigr) dt\,dr
    + \ell^2 dr^2 \,,
\end{align}
and $h^{\mu \nu} \partial_\mu \partial_\nu = \frac{1}{\ell^2} \partial_r^2$ which are required to satisfy the following completeness relation
\begin{align}
    h^{\mu\nu}\tau_\nu = 0 \,, \qquad
    h_{\mu\nu}v^\nu = 0 \,, \qquad
    h^{\mu\rho} h_{\rho\nu} = \delta^\mu{}_\nu - v^\mu \tau_\nu \,.
\end{align}
This completes the Newton--Cartan data associated with the bulk gauge connection. 
In the next subsection, we show that this geometry realises the EdS-G algebra as its isometry algebra.

\subsection{Isometries of the Newton--Cartan geometry} \label{sec:NC}
In this subsection we restrict the geometry to constant values of $\mathcal{L}(t) = \mathcal{L}_0$ and $\mathcal{T}(t) = \mathcal{T}_0$, for simplicity. The Newton--Cartan geometry we consider, therefore, takes the form 
\begin{align} \label{eq:nc-geom}
    \tau_\mu dx^\mu &= \ell \mathcal{T}_0\, dt \,, \\
    e_\mu dx^\mu &= \ell \Bigl(1 + \mathcal{L}_0 + r \mathcal{T}_0\Bigr) dt - \ell\, dr \,, \\
    m_\mu dx^\mu &= 2 \ell r \mathcal{L}_0\, dt \,.
\end{align}
What makes the above truly a Newton--Cartan geometry is the fact that, in addition to the Galilean boost-invariant quantities $h^{\mu\nu}$ and $\tau_\mu$, one can construct the following boost-invariant combinations \cite{Bergshoeff:2014uea, Hartong:2014oma}
\begin{align}\label{eq:invariant}
    \hat{v}^{\mu}
    &= 
    v^{\mu}
    - 
    h^{\mu\nu} m_{\nu}
    \,, \\
    \hat{h}_{\mu\nu}
    &= 
    h_{\mu\nu}
    -
    \tau_{\mu} m_{\nu}
    -
    \tau_{\nu} m_{\mu}
    \,.
\end{align}
The presence of the gauge field $m_\mu$ captures the Bargmann symmetry of this two-dimensional model. Let us emphasise this point clearly by examining the isometries. Given the Killing vector $\xi$, the objects $\{h^{\mu \nu}, \tau_\mu \}$ transform as usual spacetime tensors by a Lie derivative, namely
\begin{align} \label{eq:Lie-deriv-simp}
       \mathcal{L}_\xi h^{\mu \nu} = 0\,, \hspace{0.5cm} \mathcal{L}_\xi \tau_\mu = 0\,.
\end{align}
Notice that these objects of the Newton-Cartan geometry are Galilean boost-invariant, so their transformation laws do not exhibit any gauge parameter. For the rest of the tensors built upon from these, it is possible to group both diffeomorphisms $\xi$ and gauge symmetries $\lambda$ in one symmetry transformation $\delta_{\epsilon}\equiv \mathcal{L}_\xi+\delta_{\lambda}$, generating the following symmetry transformations for the remaining tensors
\begin{align} \label{eq:Lie-deriv-var}
    \mathcal{L}_\xi v^\mu  = h^{\mu \nu} \lambda_\nu\,, \hspace{0.5cm} \mathcal{L}_\xi h_{\mu \nu} = 2 \lambda_{(\mu} \tau_{\nu )}\, \hspace{0.5cm} \mathcal{L}_\xi m_\mu = \partial_\mu \sigma + \lambda_\mu \,,
\end{align}
where $\sigma$ parametrises the $U(1)$ gauge parameter of $m_\mu$ while $\lambda_\mu$ denotes the Galilei boost parameters. Notice that we used the completeness relations \eqref{eq:completenesseq} to derive these expressions and they also show that the combinations introduced in \eqref{eq:invariant} are indeed boost invariant. 

Let us now turn to working out the symmetries. Given the Newton-Cartan geometry in \eqref{eq:nc-geom}, we insert the ansatz 
\begin{align}
    \xi = \xi^t(t,r)\partial_t + \xi^r(t,r)\partial_r\,, 
    \qquad 
    \lambda_\mu = \lambda_\mu(t,r)\,,
    \qquad
    \sigma = \sigma(t,r)\,,
\end{align}
into \eqref{eq:Lie-deriv-simp}–\eqref{eq:Lie-deriv-var} and find a coupled system of partial differential equations. The first two  constraints, $\mathcal{L}_\xi h^{\mu\nu}=0$ and $\mathcal{L}_\xi \tau_\mu = 0$ from \eqref{eq:Lie-deriv-simp} together imply that
\begin{align}
    \partial_r \xi^t = 0\,,
    \qquad
    \partial_r \xi^r = 0\,, \qquad \partial_t \xi^t = 0\,.
\end{align}
so these already imply that $\xi^t = \mathrm{const}$ and $\xi^r = \xi^r(t)$, while the remaining conditions in \eqref{eq:Lie-deriv-var} determine the gauge parameters $\lambda_\mu$ and $\sigma$ in terms of $\xi^\mu$. Solving the full system leads to
\begin{equation}
\begin{aligned}
    \xi &= \frac{b_0}{\mathcal{T}_0} \partial_t + \Big( b_1 \cosh(\mathcal{T}_0 t) + b_2 \sinh(\mathcal{T}_0t) \Big)\, \partial_r \,,  \\
    \lambda &= \ell \left( b_1 - b_2 \right)(\cosh(\mathcal{T}_0 t) - \sinh(\mathcal{T}_0 t)) \bigg( \left( \mathcal{T}_0 r + \mathcal{L}_0 + 1  \right) \partial_t - \partial_r \bigg)  \,, \\
    \sigma &= b_3 + \frac{\ell}{\mathcal{T}_0} \bigg((b_1+b_2)\mathcal{L}_0 e^{\mathcal{T}_0 t} + (b_1-b_2)(\mathcal{T}_0 r + 1) e^{-\mathcal{T}_0 t}  \bigg) \,,
\end{aligned}
\end{equation}
where $b_0,b_1,b_2,b_3$ are integration constants. To define the extended algebra we take the tuples $(\xi_i, \sigma_i)$ for $i = 0,1,2,3$ where for each $i$ we set all $b_j = 0$ except $b_i$. Now, we define the commutator by
\begin{align} \label{eq:generalized-commutator-def}
     \comm{(\xi_i, \sigma_i)}{(\xi_j, \sigma_j)} := \bigg(\comm{\xi_i}{\xi_j}, \xi_i(\sigma_j) - \xi_j(\sigma_i)\bigg)\,.
 \end{align}
 This bracket arises from combining diffeomorphisms with $U(1)$ gauge transformations and corresponds to the semi-direct product structure of spacetime and the Bargmann symmetries. Thus, we find (with some suitable rescaling of the constants $b_i$) that
\begin{equation}
\begin{aligned}  \label{eq:GdSkilling_diffeo}
    H &= (\xi_0, \sigma_0) = \left(\frac{1}{\ell \mathcal{T}_0} \partial_t, 0 \right)\,, \\
    G &= (\xi_1, \sigma_1) = \left(-2 \ell \cosh(\mathcal{T}_0 t) \partial_r,\,
    -\frac{2\ell^2}{\mathcal{T}_0} \left[ \mathcal{L}_0 e^{\mathcal{T}_0 t} + (1+r \mathcal{T}_0)e^{-\mathcal{T}_0 t}
    \right] \right)\,, \\
    P &= (\xi_2, \sigma_2) = \left(2 \sinh(\mathcal{T}_0 t) \partial_r,\,
    \frac{2\ell}{\mathcal{T}_0} \left[ \mathcal{L}_0 e^{\mathcal{T}_0 t} -(1+r \mathcal{T}_0)e^{-\mathcal{T}_0 t}
    \right] \right)\,, \\
    M &= (\xi_3, \sigma_3) = (0, 4\ell^2 )\,.
\end{aligned}
\end{equation}
Applying the definition in \eqref{eq:generalized-commutator-def} we see that the generators satisfy the EdS-G algebra given in \eqref{eq:GaldSalgebra} when identifying $\tilde{\Lambda}=1/\ell^{2}$.

\subsection{Relativistic uplift}\label{sec:uplifting}

A standard way to relate Newton--Cartan geometry to a higher-dimensional relativistic spacetime is through a null uplift \cite{Duval:1984cj,Hartong:2022lsy}. This provides a useful dictionary between the two-dimensional Newton--Cartan data and higher-dimensional Lorentzian quantities. Concretely, the uplift is defined by the metric ansatz
\begin{align}
    ds^2 = h_{\mu \nu} dx^\mu dx^\nu + 2 \tau_\mu dx^\mu (du - m_\nu dx^\nu) \,,
\end{align}
where $u$ is a newly introduced null coordinate. Restricting the Newton--Cartan geometry in \eqref{eq:nc-geom} to the constant sector $\mathcal{L}(t)=\mathcal{L}_0$ and $\mathcal{T}(t)=\mathcal{T}_0$, a direct computation shows that the uplifted three-dimensional metric is Ricci flat and takes the form
\begin{align} \label{eq:uplift-metric}
    ds^2 &= \ell^2 \Big[(1+\mathcal{L}_0 + r\mathcal{T}_0)^2 - 4 r \mathcal{L}_0 \mathcal{T}_0\Big] dt^2 - 2\ell^2 (1+\mathcal{L}_0 + r\mathcal{T}_0)\, dt\,dr 
    + \ell^2 dr^2 
    + 2\ell \mathcal{T}_0\, dt\,du \,.
\end{align}
It is straightforward to verify that $\partial_u$ is null with respect to \eqref{eq:uplift-metric} and that all metric components are independent of $u$. The uplift therefore takes the standard Bargmann form associated with null uplifts of Newton--Cartan geometry \cite{Duval:1984cj,Hartong:2022lsy}.
Considering the three-dimensional Lorentzian geometry, we find the following Killing vectors 
\begin{equation}
\begin{aligned} \label{eq:GdS3d}
    H &= \frac{1}{\ell \mathcal{T}_0}\, \partial_t \,, \\[0.2cm]
    G &= -2\ell \cosh \left(\mathcal{T}_0 t \right)\, \partial_r 
    - \frac{2\ell^2}{\mathcal{T}_0} \bigg(
        \mathcal{L}_0 e^{\mathcal{T}_0 t}
        + (1 + r\mathcal{T}_0)e^{-\mathcal{T}_0 t}
    \bigg) \partial_u \,, \\[0.2cm]
    P &= 2\ell \sinh(\mathcal{T}_0 t)\, \partial_r 
    + \frac{2\ell^2}{\mathcal{T}_0} \bigg(
        \mathcal{L}_0 e^{\mathcal{T}_0 t}
        - (1 + r\mathcal{T}_0)e^{-\mathcal{T}_0 t}
    \bigg) \partial_u \,, \\[0.2cm]
    M &= 4\ell^2\, \partial_u \,,
\end{aligned}
\end{equation}
which we again recognise as the EdS-G generators from equation  \eqref{eq:GaldSalgebra}. Comparing \eqref{eq:GdS3d} with the two-dimensional Newton--Cartan generators in \eqref{eq:GdSkilling_diffeo}, one sees that the higher-dimensional isometries decompose into ordinary spacetime diffeomorphisms on $(t,r)$ together with translations along the null direction $u$. From the lower-dimensional viewpoint, the latter are precisely the $U(1)$ gauge transformations of the mass gauge field $m_\mu$.

\subsection{Non-relativistic Jackiw-Teitelboim gravity from an action principle} \label{sec:NR-JT}
Having identified the bulk Newton--Cartan geometry associated with the gauge connection, we now ask whether the same structure can be obtained from a second-order action principle. In this subsection, we show that this is indeed the case: the geometry constructed above arises as a solution of a non-relativistic JT-type model formulated directly in Newton--Cartan variables. This provides a complementary second-order description of the bulk theory and sets the stage for the comparison with the first-order gauge-theoretic formulation in Section \ref{sec:firstorder}. The corresponding action, proposed in \cite{Gomis:2020wxp}, takes the form\footnote{An equivalent form of the action, which is simply a  rewriting of \eqref{eq:NR-JT} in terms of one-form variables, is
\begin{align}
I[\tau,e,m,\phi] = \kappa \int d^2x \,\epsilon^{\mu\nu}\phi\left(\partial_\mu\omega_\nu - \frac{1}{\ell^2}\tau_\mu e_\nu\right) \,.
\end{align}
The equations of motion for this equivalent formulation are presented in appendix \ref{app:NRJT-eom-lowered}.}
\begin{align} \label{eq:NR-JT}
    I_{\mathrm{NR-JT}}
    =
    \kappa \int_{\mathcal{M}} d^2x \, E \, \phi \left( R^{\mathrm{NR}} - \frac{2}{\ell^2} \right) \,.
\end{align}
where the measure form is $E = -\epsilon^{\mu \nu} \tau_\mu e_\nu$ and the non-relativistic Ricci scalar and spin connection are defined as
\begin{align} \label{eq:RNRandspin}
    R^{\mathrm{NR}} = 4 v^{[\mu} e^{\nu]} \partial_\mu \omega_\nu\,, \hspace{1cm} \omega_\nu = 2 v^{[\alpha} e^{\beta]} \left( \partial_\alpha(e_\beta) e_\nu - \partial_\alpha(m_\beta) \tau_\nu \right)\,.
\end{align}
The action contains four dynamical fields $\{ \tau_\mu, e_\mu, m_\mu, \phi\}$ with the spin-connection $\omega_\mu$ depending on them and therefore not being a dynamical field in this formalism. The expression of the spin-connection comes from solving the field equations in the first-order formalism algebraically. Moreover, this NR-JT action \eqref{eq:NR-JT} can be understood as arising from a non-relativistic $\frac{1}{c^2}$ expansion of relativistic JT gravity, in close analogy with the non-relativistic expansion of general relativity studied in \cite{Hansen:2020pqs}. 
In the present formulation, the expression for $R^{\mathrm{NR}}$ should be viewed as a convenient scalar capturing the geometric structure of the theory, rather than as the Ricci scalar constructed from a standard non-relativistic connection. 
In particular, we do not attempt to express $R^{\mathrm{NR}}$ in terms of a curvature scalar constructed from a non-relativistic connection such as the torsionless Newton--Cartan connection compatible with $\tau_\mu$ and $h^{\mu\nu}$ (see e.g. \cite{Hartong:2022lsy}). We leave this for future work.

Let us also remark that the action is invariant under a Galilean-de Sitter boost given 
\ba
\delta_{\rm EdS-G}\phi = 0\,,  \hspace{0.5cm} 
\delta_{\rm EdS-G} \tau_\mu  = 0\,, \hspace{0.5cm}
\delta_{\rm EdS-G} e_\mu =\lambda_G \tau_\mu\,,\hspace{0.5cm}
\delta_{\rm EdS-G} m_\mu  = \lambda_G e_\mu\,,
\ea 
with $\lambda_G=\lambda_{G}(t,r)$ the Galilean gauge parameter boost. We used the condition $d\tau=0$ to prove this invariance.

Varying the action with respect to $\phi$, $\tau_\mu$, $e_\mu$, and $m_\mu$ yields the equations of motion
\begin{align} \label{eq:NRJT-EOM}
    \delta \phi:& \hspace{1cm} R^{\mathrm{NR}} - \frac{2}{\ell^2} = 0 \,, \\
    \delta m_\mu:& \hspace{1cm} \partial_\alpha \left( \tau^{[\alpha}e^{\mu]}\tau_{\nu} \partial_{\beta}(E\, \phi \,\tau^{[\beta} e^{\nu]})  \right) = 0\,, \\
    \delta \tau_\mu:& \hspace{1cm} E\, \phi R^{\mathrm{NR}} \tau^\mu = \partial_\rho \left( 4 E \phi \tau^{[\rho} e^{\nu]}\right)\omega_\nu \tau^\mu + \tau^{[\alpha}e^{\beta]} \partial_\alpha(m_\beta)\partial_\nu\left( 8 E \phi \tau^{[\nu} e^{\mu]} \right)\,,\label{eq:eqw.r.t.tau} \\
    \delta e_\mu:& \hspace{1cm} E \phi R^{\mathrm{NR}} e^\mu = \partial_\rho \left( 4 E \phi \tau^{[\rho} e^{\nu]} \right) \omega_\nu e^\mu + \partial_\alpha \left( \tau^{[\alpha} e^{\mu]} e_\nu \partial_\beta \left( 8 E \phi \tau^{[\beta} e^{\nu]} \right) \right) \\&\hspace{4cm}+ \partial_\nu\left( 8 E \phi \tau^{[\mu} e^{\nu]} \right) \tau^{[\alpha}e^{\beta]} \partial_\alpha(e_\beta)\,, \nonumber
\end{align}
where in order to derive the equations of motion we made use of the orthogonality and completeness relations of the vielbein given in \eqref{eq:completenesseq}. In particular, starting from $\tau_\mu \tau^\nu + e_\mu e^\nu = \delta_{\mu}^{\nu}$ we find after contracting with $\tau^\mu$ and $e^\mu$, the following identities for the variations \cite{Andringa:2016sot}
\begin{align}
    \delta v^\nu
    &=
    - v^\mu v^\nu \delta \tau_\mu
    -
    e^\nu v^\mu \delta e_\mu \,, \\
    \delta e^\nu
    &=
    - v^\nu e^\mu \delta \tau_\mu
    -
    e^\mu e^\nu \delta e_\mu \,,
\end{align}
along with the variation of the spin current, given by
\begin{align}
    \delta \omega_\nu = -\omega_\nu \tau^\rho \delta \tau_\rho + \omega_\nu e_\rho \delta e^\rho + 2 \tau^{[\alpha}e^{\beta]}\left( \partial_\alpha (\delta e_\beta) e_\nu + \partial_\alpha(e_\beta) \delta e_\nu - \partial_\alpha(m_\beta)\delta \tau_\nu - \partial_\alpha(\delta m_\beta)\tau_\nu \right)\,.
\end{align}
It is instructive to verify that the Newton--Cartan geometry in \eqref{eq:nc-geom}, obtained previously from the gauge connection, also solves the equations of motion above, precisely when constraining to the saddle point values we found in \eqref{eq:saddle-sol}. 
\subsection{The relationship between the first and second order formulations} \label{sec:firstorder}
Having shown that the Newton--Cartan geometry introduced above arises from a second-order non-relativistic JT-type action, we now compare this description with the first-order gauge theory formulation underlying the boundary construction. Our goal is to show how the bulk fields appearing in the Newton--Cartan description are related to the variables of the first-order formalism, and to clarify in what sense the second-order equations reproduce only part of the dynamics of the full first-order theory.

The BF action principle for Galilean de Sitter gravity in the first-order formalism is \cite{Gomis:2020wxp}
\begin{align} \label{eq:action-first-order}
   I_{\mathrm{BF}} = c_1 \int_{\mathcal{M}} \left[ \phi R(G) + \frac{1}{\ell^2}\bigg( \eta R(M) + \zeta R(H) - \rho R(P) \bigg) \right] + \frac{c_0}{\ell^2} \int_{ \mathcal{M}} \eta R(H)\,.
\end{align}
Here $c_0$ and $c_1$ are the two arbitrary constants parametrising the invariant bilinear form in \eqref{eq:pairing}, expressed in the Galilean basis and the two-form curvatures are given by
\begin{align} \label{eq:curvatures-first-order}
    R(H) = d\tau \,, \hspace{0.5cm}
    R(P) = de + \omega \tau \,, \hspace{0.5cm}
    R(G) = d\omega - \frac{1}{\ell^2}\tau e \,, \hspace{0.5cm}
    R(M) = dm + \omega e \,.
\end{align}
The fields $\{\eta,\zeta,\rho,\phi\}$ are scalars, while
$\{\tau,e,\omega,m\}$ are one-forms. In this formalism, $\omega_\mu$ is still a dynamical field. A general variation of \eqref{eq:action-first-order} produces the boundary term
\begin{align}
    \Theta = c_1 \int_{\partial \mathcal M} \Bigg(  \phi \delta \omega  + \frac{1}{\ell^2} \Big(\eta \delta m+\zeta \delta \tau-\rho \delta e  \Big)\Bigg) + \frac{c_0}{\ell^2} \int_{\partial \mathcal{M}} \eta \delta \tau\,.
\end{align}
To obtain a well-defined variational principle, we impose boundary conditions at the asymptotic de Sitter boundary, such that the scalar fields are proportional to the temporal components of the gauge fields,
\begin{align}\label{bndrycond}
    \eta = \tau_t\,, \hspace{0.5cm} 
\rho = e_t\,, \hspace{0.5cm}
\phi = \omega_t \,, \hspace{0.5cm}
\zeta = m_t\,, 
\end{align}
 We conclude that the action must then in addition to \eqref{eq:action-first-order} be supplemented by the boundary term
 \begin{align} \label{eq:bdry-term-gds}
    I_\partial = -\frac{c_1}{2} \int dt \left( \omega_t^2 + \frac{1}{\ell^2}(2 \tau_t m_t - e_t^2) \right) - \frac{c_0}{2 \ell^2} \int dt \, \tau_t^2 \,.
 \end{align}
Now that the variational principle is well defined, variation of the action \eqref{eq:action-first-order} with respect to $\{\eta,\rho,\phi,\zeta\}$ imposes the flatness constraints
\begin{align} \label{eq:flatness}
    R(H)=0\,, \qquad R(P)=0\,, \qquad R(G)=0\,, \qquad R(M)=0\,,
\end{align}
for the curvatures defined in \eqref{eq:curvatures-first-order}. One may solve algebraically these flatness constraints in \eqref{eq:flatness} to determine the spin connection $\omega$ in terms of the other fields $m, \tau$ and $e$ finding
\begin{align} \label{eq:omega-second-order}
    \omega_\mu
    =
    \frac{1}{E}\,\epsilon^{\alpha\beta}
    \bigl(
        \partial_\alpha (m_\beta)\, \tau_\mu
        -
        \partial_\alpha (e_\beta)\, e_\mu
    \bigr)\,,
\end{align}
where $E = -\epsilon^{\mu\nu}\tau_\mu e_\nu$. Note that this was precisely the form we used to define the second-order formulation in \eqref{eq:RNRandspin}. The variation with respect to $\{\tau,e,\omega,m\}$ gives
\begin{align}\label{scalareq}
    d\eta  = 0\,, \hspace{0.4cm}
d\rho + \omega \eta - \tau \phi  =0\,, \hspace{0.4cm} 
d\phi -\frac{1}{\ell^2} (\tau \rho -e\eta)=0\,,  \hspace{0.4cm}
d\zeta + \omega \rho - e \phi =0\,.
\end{align}
Solving the curvature constraints \eqref{eq:curvatures-first-order} subject to the boundary conditions \eqref{bndrycond}, we obtain
\bseq \label{solution}	
\ba
\tau &=& \ell \mathcal{T}(t) \, dt\,, \\
e &= & \ell \big( -dr + \left( r\mathcal{T}(t) + 1+ \mathcal{L}(t)\right)\, dt\big)\,,\\
\omega &= &  -dr + \big( r\mathcal{T}(t) + 1- \mathcal{L}(t)\big)\, dt\,,\\
m &=&2\ell r \mathcal{L}(t) dt\,.
\ea
\eseq
where $\mathcal{T}(t)$ and $\mathcal{L}(t)$ are for now arbitrary functions. Substituting the temporal components of the gauge fields in \eqref{solution} into the boundary term \eqref{eq:bdry-term-gds}, and using the expressions for $\mathcal{T}$ and $\mathcal{L}$ in \eqref{eq:LandT}, we recover the Schwarzian-like action \eqref{eq:Galilean-de-Sitter}.
Furthermore, the scalar equations imply additional constraints on the auxiliary fields. From $d\eta=0$ we obtain $\eta = \eta_0$ with $\eta_0$ constant. Using the solution \eqref{solution} in \eqref{scalareq}, one finds
\begin{align}
    \phi(t,r) = \frac{\eta_0}{\ell}\,r+\phi_0(t)\,, \hspace{0.5cm}
    \rho(t,r) = \eta_0\,r+\rho_0(t)\,,
\end{align}
for arbitrary functions $\phi_0(t)$ and $\rho_0(t)$. Substituting these expressions back into \eqref{scalareq} then yields
\begin{align}
    \eta_0 \bigl(1-\mathcal{L}(t)\bigr) - \ell \mathcal{T}(t)\phi_0(t) + \frac{d\rho_0(t)}{dt} &= 0\,, \\
    \eta_0 \bigl(1+\mathcal{L}(t)\bigr) - \mathcal{T}(t)\rho_0(t) + \ell \frac{d\phi_0(t)}{dt} &= 0\,.
\end{align}
Combining the last two equations, we obtain
\be\label{phi2}
\Bigg( \frac{d^2}{dt^2} -\frac{d\big(\log(\mathcal{T})\big)}{dt}\,\frac{d}{dt}+V\big(\phi_0, \mathcal L,\mathcal T\big)\Bigg)\, \phi_0(t)=0\,, 
\ee
with the potential 
\be
V\big(\phi_0, \mathcal L,\mathcal T\big)= \frac{\eta_0}{\ell} \Bigg(\mathcal T\big(1- \mathcal L\big) +\frac{d\mathcal{L} }{dt}-\frac{d\big(\log(\mathcal T)\big)}{dt} -\mathcal{L}\, \frac{d\big(\log(\mathcal T)\big)}{dt}\Bigg)- \mathcal{T}^2\, \phi_0\,.
\ee
We emphasise that neither $\rho$ nor $\eta$ appears in the second-order action \eqref{eq:NR-JT}. A comparison with the first-order formalism suggests that the second-order description corresponds to a partially reduced sector of the full first-order theory, in which part of the auxiliary scalar structure has already been solved for or fixed. In particular, the first-order scalar fields $\rho$ and $\eta$ do not appear as independent variables in the second-order action, indicating that their dynamics are encoded only implicitly through the reduced variables.
We emphasise that the equation \eqref{phi2} is reproduced in the second-order formalism, and therefore provides a kinematical equation for the scalar field $\phi$. However, it is subject to the constraining equation \eqref{eq:eqw.r.t.tau} coming from variation with respect to $\tau_\mu$ in the second-order formalism, suggesting that the gauge field $\tau_\mu$ could be interpreted as a Lagrange multiplier in this formalism.\footnote{We thank Professor Jorge Zanelli for pointing this suggestion out.}

\section{Discussion}\label{sec:discuss}
In this paper, we considered the two-dimensional Galilean version of de Sitter space, of which the symmetries are captured by the extended de Sitter-Galilei (EdS-G) algebra. We studied aspects of both its bulk and boundary realisation. 

On the boundary side, we employed a Schwarzian-like action to compute one-loop contributions. We report that these contributions capture quantum fluctuations that dominate the low-energy density of states and at low temperatures we found the partition function to be domintated by $\mathcal{Z}_{\text{EdS-G}}(\beta)\sim \beta^{-2}$. Here this $2$ correctly reflects the factors of $1/2$ that are added corresponding to the four generators of the EdS-G algebra. To arrive at this result, we derived the symplectic form entering the path integral measure directly from the effective action by introducing Ostrogradsky canonical momenta. To the best of our knowledge, this action-based derivation of the symplectic form has not been emphasised in the literature. We also found that the logarithmic contribution is not accompanied by an imaginary phase, in agreement with corresponding two-dimensional relativistic computations \cite{Maldacena:2019cbz,Cotler:2019nbi}. This contrasts with the relativistic result from four dimensions \cite{Blacker:2025zca,Maulik:2025phe}, which may be attributed to the intrinsically two-dimensional nature of the present computation. 
It should also be pointed out that central charge $c_{1}$ drops out of our final result -- this is potentially a side-effect of choosing our vacuum in \eqref{eq:saddle-sol}. We furthermore assumed, in our computation, that the answer is one-loop exact. In light of the close relation to the warped Schwarzian theory \cite{Afshar:2019tvp}, this appears plausible, although it certainly warrants further investigation. An exciting direction for future work would be to construct a SYK-like realisation of this model. A possible source of inspiration in this direction is \cite{Afshar:2019axx}, where related ideas were explored in the context of flat space holography and the complex SYK model and its corresponding Carrollian counterpart.

On the bulk side we show that once the EdS-G algebra is geometrised into a Newton-Cartan language, that one can construct a JT-like action of which this geometry solves the equations of motion. We note in particular that, upon restricting to the torsionless sector with $d\tau=0$, the equation obtained by varying the JT-like action with respect to $\tau_\mu$ involves only first derivatives. It would be interesting to understand whether this JT-like action reflects a deeper Lagrange-multiplier role for $\tau_\mu$, or whether it is simply a feature of the torsionless truncation. We also leave a careful Hamiltonian analysis of this action for future discussion. 
A natural next step for future study is to study bulk fluctuations around the here studied background, along lines similar to recent analyses of logarithmic contributions and Lichnerowicz-type zero modes in near-extremal black holes \cite{Maulik:2024dwq, Blacker:2025zca, Maulik:2025phe, PandoZayas:2026vbg}. The goal would be to use these zero modes to verify our prediction regarding the exponent factor of $\beta^{-2}$ entering in the partition function. Finally, we think this setup may also aid the exploration of studying thermal gravitational settings in a non-Lorentzian setting.

We want to end by emphasising that in this paper we have contributed towards paving the way for extending the holographic duality in low dimensions beyond relativistic symmetries. In conjunction we also believe our results contribute to shaping tools for further unwrapping the mysteries of de Sitter spacetime.
\subsection*{Acknowledgments}
We thank Hamid Afshar, Patricio-Salgado Rebolledo, and Jorge Zanelli for fruitful comments. 
MH is supported by the Icelandic Research Fund under grant 228952-053 and the University of Iceland Doctoral Grant HEI2025-96517. DH is supported by the Icelandic Research Fund 2511228-051. DH also thanks the Institute for Theoretical Physics of Utrecht University for its hospitality, where part of the project was carried out, and, in particular, Professor Stefan Vandoren and Huaxuan Zeng for fruitful discussions on related topics. MH and DH also thank the Nordita Institute for the kind hospitality in which part of this project was undertaken. MH thanks Chalmers tekniska högskola for the kind hospitality in which part of this project was undertaken. The work of WS is supported by Starting Grant 2023-03373 from the Swedish Research Council and the Olle Engkvists Stiftelse.

\appendix

\section{Ostrogradsky symplectic form for the Schwarzian theory} \label{sec:Ostro-sch}

In this appendix, we show that the symplectic structure of the Schwarzian theory \cite{Witten:1987ty, Alekseev:1988ce, Stanford:2017thb, Saad:2019lba} can be recovered directly from Ostrogradsky variables.

We begin with the Schwarzian theory
\begin{align}
I_{\mathrm{Sch}}[f]
= - \int d\tau \left(  \mathrm{Sch}(f,\tau) + \frac{1}{2}{f'}^2 \right) \,.
\end{align}
We now expand to second order with $f = \bar{f} + \delta f$ deriving the 
\begin{align}
    I^{(1)}_{\mathrm{Sch}} &= \int d\tau \,  \partial_\tau\left( \bar{f}' - \frac{\bar{f}''^2}{\bar{f}'^3} + \frac{\bar{f}'''}{\bar{f}'^2}\right)\delta f \,, \\
    I^{(2)}_{\mathrm{Sch}} &= \int \frac{d\tau}{\bar{f}'^2} \left[ -\frac{1}{2}\bar{f}'^2 \delta f'^2 + \frac{3}{2} \frac{\bar{f}''^2}{\bar{f}'^2} \delta f'^2 - 2 \frac{\bar{f}''}{\bar{f}'} \delta f' \delta f'' + \frac{1}{2}\delta f''^2 \right]\,.
\end{align}
Now, we compute the Ostrogradsky conjugate momenta to $\delta f$ to find
\begin{align}
    p_{\delta f} = \frac{\partial L^{(2)}_{\mathrm{Sch}}}{\partial (\delta f')} - \frac{d}{d\tau}\left( \frac{\partial L^{(2)}_{\mathrm{Sch}}}{\partial( \delta f'')} \right) = \frac{2}{\bar{f'}^2}\left[ \mathrm{Sch}\left( \tanh \tfrac{\bar{f}}{2}, \tau \right) \delta f' + \tfrac{\bar f''}{{\bar f'}}\delta f'' - \tfrac{1}{2} \delta f'''\right]\,.
\end{align}
Hence, the symplectic form which we will use to compute the measure is given by
\begin{align}
    \Omega_{\mathrm{Sch}}^{(2)}
    &= \int_0^\beta d\tau \, dP_{\delta f} \wedge d(\delta f)
    \nonumber\\
    &= \int_0^\beta \frac{2\,d\tau}{\bar f'^2}\left[
        \mathrm{Sch}\!\left(\tanh\frac{\bar f}{2},\tau\right)
        d\delta f' \wedge d\delta f
        + \frac{\bar f''}{\bar f'} d\delta f'' \wedge d\delta f
        - \frac{1}{2} d\delta f''' \wedge d\delta f
    \right].
\end{align}
The last term can be integrated by parts to find
\begin{align}
  \Omega^{(2)}_{\mathrm{Sch}} =   \int_0^\beta \frac{d\tau}{\bar f'^2}\left[ d\delta f'' \wedge d\delta f' + 2\, \mathrm{Sch}(\tanh\tfrac{\bar f}{2},\tau) d\delta f' \wedge d\delta f \right]\,.
\end{align}
Once evaluated on the classical solution $\bar{f}(\tau) = \frac{2\pi}{\beta}\tau$ this gives the same as equation (114) of \cite{Saad:2019lba} and equation (2.8) of \cite{Stanford:2017thb}.

\section{One-loop analysis for the warped Schwarzian action} \label{sec:Ostro-warped-sch}

In this appendix, we revisit the quadratic fluctuations of the warped Schwarzian action of \cite{Afshar:2019tvp}. Our aim is to show that the one-loop partition function can be obtained directly from the Ostrogradsky symplectic form, without appealing to the coadjoint orbit formulation. Starting from equation (5.5) of \cite{Afshar:2019tvp}, we expand the fields in Fourier modes and rewrite the result as a sum over positive modes only. This gives
\begin{align}
    S^{(2)} &= \int_0^\beta d\tau \left[ \frac{c}{24} \left( {\epsilon''}^2 - \left( \frac{2\pi}{\beta}\right)^2 {\epsilon'}^2\right) + \frac{k}{4} \left(\sigma' + \alpha \epsilon'  \right)^2 - \kappa \left( \epsilon''  + \frac{2\pi i}{\beta}\epsilon' \right) \left(\sigma' + \alpha \epsilon' \right) \right] \nonumber \\
    &= \frac{4\pi^2}{\beta} \sum_n \left[ \frac{c}{24} \left( \frac{2\pi}{\beta} \right)^2 n^2 (n^2-1) \epsilon_n \epsilon_{-n} + \frac{k}{4} n^2 \tilde{\sigma}_n \tilde{\sigma}_{-n} - \frac{2\pi i}{\beta} \kappa n^2 (n+1) \epsilon_n \tilde{\sigma}_{-n} \right] \nonumber \\
    &= \frac{8 \pi^2}{\beta} \sum_{n \geq 2} \frac{c}{24}\left( \frac{2\pi}{\beta} \right)^2 n^2(n^2-1)\abs{\epsilon_n}^2 - \frac{8 \pi^2}{\beta}i \sum_{n \geq 1} \frac{2\pi i}{\beta} \kappa n^3 \left( \epsilon_n^{(I)} \tilde{\sigma}_n^{(R)} - \epsilon_n^{(R)} \tilde{\sigma}_n^{(I)} \right) \nonumber  \\
    &\hspace{1cm}  + \frac{8 \pi^2}{\beta} \sum_{n \geq 1} \frac{k}{4} n^2 \abs{\tilde{\sigma}_n}^2 - \frac{8 \pi^2}{\beta}\sum_{n \geq 1} \frac{2\pi i}{\beta} \kappa n^2 \left( \epsilon_n^{(R)} \tilde{\sigma}_n^{(R)} + \epsilon_n^{(I)}\tilde{\sigma}_n^{(I)} \right)\,.
\end{align}
We introduced $\tilde{\sigma} = \sigma + \alpha \epsilon$ to simplify the resulting expressions. Note that the \(n=1\) mode is degenerate, since the determinant of the corresponding quadratic form vanishes. This sector must therefore be treated separately in the path integral. We now turn to the symplectic structure associated with the quadratic action. Using the Ostrogradsky prescription, the quadratic symplectic form is constructed, where the canonical momenta are given by
\begin{align}
    p_\epsilon &= \frac{\partial \mathcal{L}^{(2)}}{\partial \epsilon'} - \frac{d}{d\tau}\left(\frac{\partial \mathcal{L}^{(2)}}{\partial \epsilon''} \right) = \frac{1}{2}k \alpha \tilde{\sigma}' + \kappa \sigma'' - \frac{c \pi^2}{3 \beta^2}\epsilon' - \frac{4\pi i \alpha \kappa}{\beta} \epsilon' - \frac{2\pi i}{\beta} \kappa \sigma' - \frac{1}{12}c \epsilon''' \,, \\
    p_\sigma &= \frac{\partial \mathcal{L}^{(2)}}{\partial \sigma'} - \frac{d}{d\tau}\left(\frac{\partial \mathcal{L}^{(2)}}{\partial \sigma''} \right)  = \frac{1}{2}k \tilde{\sigma}' - \kappa \epsilon'' - \frac{2\pi i}{\beta} \kappa \epsilon' \,.
\end{align}
Therefore we find 
\begin{align}
    \Omega^{(2)}
    &=
    \int_0^\beta d\tau \Bigg[
        \tfrac{1}{2}k\, d\tilde{\sigma}' \wedge d\tilde{\sigma}
        - \tfrac{c}{12}\Bigl(
            d\epsilon''' \wedge d\epsilon
            + \left(\tfrac{2\pi}{\beta}\right)^2 d\epsilon' \wedge d\epsilon
        \Bigr)
        + 2\kappa\, d\sigma' \wedge d\epsilon
        - \tfrac{4\pi i}{\beta}\kappa\, d\tilde{\sigma}' \wedge d\epsilon
    \Bigg] \notag \\
    &=
    2\pi i \sum_n \Bigg[
        \tfrac{1}{2} k  n \,
        d\tilde{\sigma}_n \wedge d\tilde{\sigma}_{-n}
        + \tfrac{c}{12} \left(\tfrac{2\pi}{\beta}\right)^2 n(n^2-1)\,
        d\epsilon_n \wedge d\epsilon_{-n}
        + 2 i \kappa \left(\tfrac{2\pi}{\beta}\right) n(n-1)\,
        d\tilde{\sigma}_n \wedge d\epsilon_{-n}
    \Bigg]  \notag \\
    &=
    4\pi k \sum_{n\geq 1} n \,
    d\tilde{\sigma}_n^{(R)} \wedge d\tilde{\sigma}_n^{(I)}
    + 4 \beta \sum_{n\geq 2}
    \frac{c}{12}\left(\frac{2\pi}{\beta}\right)^3 n(n^2-1)\,
    d\epsilon_n^{(R)} \wedge d\epsilon_n^{(I)}
    \\
    &\hspace{1cm}
    - 4 \beta \kappa \sum_{n\geq 1}
     \left(\frac{2\pi}{\beta}\right)^2 n^2
    \Bigl(
        d\tilde{\sigma}_n^{(R)} \wedge d\epsilon_n^{(R)}
        +
        d\tilde{\sigma}_n^{(I)} \wedge d\epsilon_n^{(I)}
    \Bigr) \nonumber
    \\
    &\hspace{2cm}
    + 4\beta \kappa i \sum_{n\geq 1}
     \left(\frac{2\pi}{\beta}\right)^2 n
    \Bigl(
        d\tilde{\sigma}_n^{(I)} \wedge d\epsilon_n^{(R)}
        -
        d\tilde{\sigma}_n^{(R)} \wedge d\epsilon_n^{(I)}
    \Bigr) \nonumber \, .
\end{align}

With respect to the basis $x = \left\{ \tilde{\sigma}_n^{(R)}, \tilde{\sigma}_n^{(I)}, \epsilon_n^{(R)}, \epsilon_n^{(I)}  \right\}$ we may write
\begin{align}
    \Omega_n^{(2)}
    &=
    \begin{pmatrix}
        0 & A_n & -n B_n & - i B_n \\
        -A_n & 0 & i B_n & -n B_n \\
        n B_n & - i B_n & 0 & C_n \\
        i B_n & n B_n & -C_n & 0
    \end{pmatrix},
    \qquad
    {\small
    \begin{aligned}
        A_n &= 4\pi k\, n \,, \\
        B_n &= 4\kappa \beta \left(\frac{2\pi}{\beta}\right)^2 n \,, \\
        C_n &= \frac{c\beta}{3}\left(\frac{2\pi}{\beta}\right)^3 n(n^2-1)\,.
    \end{aligned}
    }
\end{align}
\begin{align}
I_n^{(2)} &=
\begin{pmatrix}
D_n & 0 & -iE_n & nE_n \\
0 & D_n & -nE_n & -iE_n \\
-iE_n & -nE_n & F_n & 0 \\
nE_n & -iE_n & 0 & F_n
\end{pmatrix},
\qquad
{\small
\begin{aligned}
D_n &= \frac{2\pi^2}{\beta} k n^2, \\
E_n &= \frac{8\pi^3}{\beta^2} k n^2, \\
F_n &= \frac{c\pi^2}{3\beta}
\left(\frac{2\pi}{\beta}\right)^2
n^2 (n^2 - 1)\,.
\end{aligned}
}
\end{align}
For modes with $n\geq 2$ the quadratic fluctuation integral is a standard Gaussian.
The case $n=1$ is special, since the corresponding integral is not Gaussian and the symplectic form is degenerate, and must therefore be treated separately. 
We denote its contribution by $\mathcal{B}_1(\beta)$. Thus the one loop partition function takes the form
\begin{align}
    Z(\beta)
    =
    \mathcal{B}_1(\beta)
    \prod_{n\geq 2}
    \mathrm{Pf}\left(\Omega_n^{(2)}\right)
    \sqrt{
        \frac{(2\pi)^4}{\det I_n^{(2)}}
    } = 2 \beta \prod_{n \geq 2} \frac{4 \beta^2}{n^2}  = \frac{1}{8\pi \beta^2}\, .
\end{align}
In the last step, we used the zeta-function regularisation, and in the preceding step, we used the following identities. For $n \geq 2$, we have
\begin{align}
    \mathrm{Pf}(\Omega_n^{(2)}) &=  \frac{32 \pi^4}{3 \beta^2}n^2(n^2-1)(ck - 24 \kappa^2) \,, \\ \sqrt{\det I_n^{(2)}} &= \frac{32 \pi^6 }{3 \beta^4}n^4(n^2-1)(ck - 24 \kappa^2) \,.
\end{align}
The remaining contribution comes from the $n=1$ mode. In this case, we evaluate
\begin{align}
    \mathcal{B}_1(\beta) = 2\pi i \left[ k \int   d\tilde{\sigma}_1 d\tilde{\sigma}_1^*  + 4 i \kappa \left( \tfrac{2\pi}{\beta} \right)^2 \int  d\epsilon_1 d\tilde{\sigma}_1^*\right]  e^{\frac{4 \pi^2 i}{\beta^2}\left[  \frac{k \beta}{2}\abs{\tilde{\sigma}_1}^2 - 4\pi i \kappa \epsilon_1^* \tilde{\sigma}_1 \right]} = 2 \beta \,.
\end{align}
The first term gives \(2\beta\), while the remaining four mixed contributions evaluate to \(\pm 4\pi i\) and cancel pairwise in the sum. 
This agrees with the result obtained in \cite{Afshar:2019tvp}.

\section{The saddle from global conformal Galilean invariance} \label{ap:global-symmetries-twised}

In this appendix, we give an alternative, algebraic characterisation of the saddle used in the one-loop analysis. The key observation is that the global conformal Galilean de Sitter algebra arises as a finite-dimensional subalgebra of the warped Virasoro algebra governing the transformations of $\mathcal L(t)$ and $\mathcal T(t)$. Requiring $\mathcal L(t)$ and $\mathcal T(t)$ to be invariant under this global subalgebra then singles out precisely the constant background used in the main text.

The global conformal Galilean-de Sitter algebra can be identified as a finite-dimensional subalgebra of the warped Virasoro algebra introduced in \eqref{eq:warped-virasoro} and given by
\bseq\label{wViralgebra}
\ba
i\, \{ L_m \,, L_n \} & =& (m-n) L_{n+m}  + c\,  (n^3-n)\, \delta_{n+m,0}\,,\\
i\, \{ L_m \,, P_n\} & =& -n\, P_{m+n}  +i\kappa \,(m^2-m) \, \delta_{m+n,0}\,,\\
i\, \{ P_n \,, P_m\} & =& 0\,.
\ea
\eseq
To describe the asymptotic dynamics of Newton-Cartan geometry in $(1+1)$-spacetime dimensions, we use the conformal Galilean-de Sitter algebra given by
\be\label{eq:confGalilean}
\left[ \mathcal H\,, \mathcal D\right] =\mathcal H\,, \hspace{1cm}
\left[ \mathcal K\,, \mathcal D\right] =-\mathcal K\,, \hspace{1cm}
\left[ \mathcal H\,, \mathcal K\right] =2\,\mathcal Z\,.
\ee
We want to show that \eqref{eq:confGalilean} is contained as a subalgebra in \eqref{wViralgebra}. In effect, looking at the set $\{L_{0}\,, L_{1}\,, P_{-1}\}$, we observe that, (using the commutation relations \eqref{wViralgebra}), they satisfy the non-vanishing Poisson bracket relations
\ba
i\, \{ L_0 \,, L_{1} \}  = - L_{1}  \,, \hspace{0.5cm}
 i\, \{ L_0 \,, P_{-1} \}  =  P_{-1}\,, \hspace{0.5cm}
 i\, \{ L_{1} \,, P_{-1} \}  = P_{0}\,.
\ea
So, promoting $i\{ \,, \} \to [\,, ]$ and making the identification $L_0 = \mathcal{D} \,, \ L_{1}= \mathcal H\,, \ P_{-1}=\mathcal K \,, \ P_0 =2\mathcal Z$, we recover \eqref{eq:confGalilean} as a subalgebra of \eqref{wViralgebra}. 

The gauge parameters $\sigma(t)$ and $\chi(t)$ admit the mode expansions
\begin{align}
    \sigma (t)= \sum_{n\in \mathbb Z} e^{\frac{2\pi i n t}{\beta}}\sigma_n\,, \hspace{0.5cm} \chi (t)= \sum_{n\in \mathbb Z} e^{\frac{2\pi i n t}{\beta}}\chi_n
\end{align}
For the global subalgebra identified above, corresponding to the charges ${L_0,L_1,P_{-1}}$, the relevant modes are
\be
\Big\{ \sigma_0\,,  e^{\frac{2\pi i t}{\beta}} \sigma_{1} \Big\} \,~~~~ \Big\{ \chi_0\,,  e^{-\frac{2\pi i t}{\beta}} \chi_{-1} \Big\}\,.
\ee
We now consider the transformation laws for $\mathcal L$ and $\mathcal T$ in \eqref{eq:transformation-laws1},
\bseq\label{transLandT}
\ba
\delta_{(\sigma,\chi)} \mathcal L & = &  \sigma \mathcal{L}' + 2\mathcal{L} \sigma' + \mathcal{T} \chi' + \chi''\,,\\
\delta_{(\sigma,\chi)} \mathcal T & =& \sigma' \mathcal T + \sigma \mathcal{T}'-\sigma''\,.
\ea
\eseq
We analyse the allowed values of $\mathcal T$ and $\mathcal L$ that satisfy conformal Galilean invariance. This provides an algebraic characterisation of the background around which the one-loop fluctuation analysis is performed. Evaluating the invariance conditions $\delta\mathcal{L}=0$ and $\delta\mathcal{T}=0$ for each of the global modes $(\sigma,\chi)$ listed on the left gives the corresponding constraints on $\mathcal{T}(t)$ and $\mathcal{L}(t)$.
\begin{equation}
\begin{alignedat}{5}
(\sigma_0,\, e^{- \frac{2\pi i t}{\beta}}\chi_{-1}):
\quad &\vline\quad&
\mathcal{T}(t) &= A,\qquad &
\mathcal{L}(t) &= -\frac{e^{- \frac{2\pi i t}{\beta}}}{\beta \sigma_0}(A\beta+2i\pi)\chi_{-1}+B\,,
\\[6pt]
(e^{\frac{2\pi i t}{\beta}}\sigma_1,\, \chi_0):
\quad &\vline\quad&
\mathcal{T}(t) &= -\frac{2\pi i}{\beta}+C\,e^{- \frac{2\pi i t}{\beta}},\qquad &
\mathcal{L}(t) &= D\,e^{- \frac{4\pi i t}{\beta}}\,,
\\[6pt]
(e^{\frac{2\pi i t}{\beta}}\sigma_1,\, e^{- \frac{2\pi i t}{\beta}}\chi_{-1}):
\quad &\vline\quad&
\mathcal{T}(t) &= -\frac{2\pi i}{\beta}+E\,e^{- \frac{2\pi i t}{\beta}},\qquad &
\mathcal{L}(t) &= -\frac{E\chi_{-1}}{\sigma_1}e^{- \frac{6\pi i t}{\beta}} + F\,e^{- \frac{4\pi i t}{\beta}}\,.
\end{alignedat}
\end{equation}
Therefore, all these conditions are compatible if $A=-2\pi i/\beta,$ $B=C=D=E=F=0$, which implies that the arbitrary functions take the form
\be\label{connd}
\mathcal{T}(t) = -\frac{2\pi i}{\beta}\,, \qquad \mathcal{L}(t) =0\,.
\ee
The conditions \eqref{connd} in terms of  the Goldstone fields $v$ and $s$ translate to
\be\label{valuesvs}
v(t)= -\frac{2i\pi t}{\beta} \,, \qquad  s(t) =0\,.
\ee
This is precisely the constant configuration for $\mathcal L$ and $\mathcal T$ selected by invariance under the global Galilean de Sitter algebra.

\section{Equations of motion for the non-relativistic JT action} \label{app:NRJT-eom-lowered}

In this appendix, we present the equations of motion for the second-order non-relativistic JT action written entirely in terms of lower-index Newton--Cartan variables. This formulation is equivalent to the one presented in the main text, but its variation is somewhat simpler since the action is expressed directly in terms of the one-forms $\tau_\mu$, $e_\mu$, and $m_\mu$. We have explicitly verified that the resulting equations of motion are equivalent to those obtained from the formulation used in the main text.

The action is

\be \label{eq:NR-JT-lowered-action}
I[\tau,e,m,\phi]   = \kappa \, \int d^2x \, \epsilon^{\mu \nu} \phi \, \Big( \partial_\mu \omega_\nu - \frac{1}{\ell^2} \tau_\mu e_\nu \Big)\,,
\ee
where the spin connection is given by 
\be\label{omega}
\omega_\mu  =\frac{1}{E} \, \epsilon^{\alpha \beta}\Big( \partial_\alpha m_\beta  \tau_\mu - \partial_\alpha e_\beta   e_\mu \Big)\,,
\ee
and $E =  -\epsilon^{\nu \rho} \tau_\nu e_\rho$. If one furthermore wishes to impose the torsionless condition $d\tau=0$, this may be implemented by introducing a Lagrange multiplier $\zeta$ and adding the term
\begin{align}
\frac{\mu}{\ell^2}\int d^2x \, \zeta \,\epsilon^{\mu\nu}\partial_\mu \tau_\nu
\end{align}
to the action. Variation with respect to $\zeta$ then imposes $\epsilon^{\mu\nu}\partial_\mu \tau_\nu = 0$, which is equivalent to $d\tau=0$. We now vary the action with respect to the dynamical fields. For simplicity, boundary terms arising from integrations by parts will be omitted.
\begin{align}
\delta \phi:& \qquad
\epsilon^{\mu \nu}
\Big(
\partial_\mu \omega_\nu - \frac{1}{\ell^2} \tau_\mu e_\nu
\Big)=0\,,
\notag
\\[0.5em]
\delta m_\beta:& \qquad
\epsilon^{\alpha \beta} \epsilon^{\mu \nu}
\partial_\alpha
\Bigg(
\frac{1}{E} \partial_\mu \phi \tau_\nu
\Bigg)=0\,,
\label{eq:lowered-eoms}
\\[0.5em]
\delta \tau_\mu:& \qquad
\Big(
\epsilon^{\alpha \nu} \epsilon^{\mu \beta} e_\beta \omega_\alpha
+
\epsilon^{\alpha \beta} \epsilon^{\mu \nu} \partial_\alpha m_\beta
\Big)\partial_\nu \phi
-
\frac{E}{\ell^2} \epsilon^{\mu \nu} e_\nu \phi
=0\,,
\notag
\\[0.5em]
\delta e_\mu:& \qquad
\Bigg(
\epsilon^{\beta \nu} \epsilon^{\alpha \mu} \tau_\alpha \omega_\beta
-
\epsilon^{\mu \nu} \epsilon^{\alpha \beta} \partial_\alpha e_\beta
\Bigg)\partial_\nu \phi
+
E \epsilon^{\beta \nu} \epsilon^{\alpha \mu}
\partial_\alpha
\Bigg(
\frac{1}{E} \partial_\nu \phi \, e_\beta
\Bigg)
+
\frac{E}{\ell^2} \epsilon^{\mu \nu} \tau_\nu \phi
=0\,.
\notag
\end{align}
These equations are equivalent to \eqref{eq:NRJT-EOM} obtained from the formulation presented in the main text.

\bibliographystyle{JHEP}
\bibliography{refds}

\providecommand{\href}[2]{#2}\begingroup\raggedright\begin{thebibliography}{10}

\bibitem{Maldacena:1997re}
J.~M. Maldacena, {\it {The Large $N$ limit of superconformal field theories and supergravity}},  {\em Adv. Theor. Math. Phys.} {\bf 2} (1998) 231--252 [\href{http://arXiv.org/abs/hep-th/9711200}{{\tt hep-th/9711200}}].

\bibitem{Teitelboim:1983ux}
C.~Teitelboim, {\it {Gravitation and Hamiltonian Structure in Two Space-Time Dimensions}},  {\em Phys. Lett. B} {\bf 126} (1983) 41--45.

\bibitem{Jackiw:1984je}
R.~Jackiw, {\it {Lower Dimensional Gravity}},  {\em Nucl. Phys. B} {\bf 252} (1985) 343--356.

\bibitem{Sachdev:1992fk}
S.~Sachdev and J.~Ye, {\it {Gapless spin fluid ground state in a random, quantum Heisenberg magnet}},  {\em Phys. Rev. Lett.} {\bf 70} (1993) 3339 [\href{http://arXiv.org/abs/cond-mat/9212030}{{\tt cond-mat/9212030}}].

\bibitem{KitaevTalks}
A.~Kitaev, ``{A simple model of quantum holography}.'' \url{http://online.kitp.ucsb.edu/online/entangled15/kitaev/}, \url{http://online.kitp.ucsb.edu/online/entangled15/kitaev2/}, 2015.
\newblock Talks at KITP, April 7, 2015 and May 27, 2015.

\bibitem{Sachdev:2010um}
S.~Sachdev, {\it {Holographic metals and the fractionalized Fermi liquid}},  {\em Phys. Rev. Lett.} {\bf 105} (2010) 151602 [\href{http://arXiv.org/abs/1006.3794}{{\tt 1006.3794}}].

\bibitem{Maldacena:2016hyu}
J.~Maldacena and D.~Stanford, {\it {Remarks on the Sachdev-Ye-Kitaev model}},  {\em Phys. Rev. D} {\bf 94} (2016), no.~10 106002 [\href{http://arXiv.org/abs/1604.07818}{{\tt 1604.07818}}].

\bibitem{Sarosi:2017ykf}
G.~S{\'a}rosi, {\it {AdS$_{2}$ holography and the SYK model}},  {\em PoS} {\bf Modave2017} (2018) 001 [\href{http://arXiv.org/abs/1711.08482}{{\tt 1711.08482}}].

\bibitem{Gu:2019jub}
Y.~Gu, A.~Kitaev, S.~Sachdev and G.~Tarnopolsky, {\it {Notes on the complex Sachdev-Ye-Kitaev model}},  {\em JHEP} {\bf 02} (2020) 157 [\href{http://arXiv.org/abs/1910.14099}{{\tt 1910.14099}}].

\bibitem{Mertens:2018fds}
T.~G. Mertens, {\it {The Schwarzian theory {\textemdash} origins}},  {\em JHEP} {\bf 05} (2018) 036 [\href{http://arXiv.org/abs/1801.09605}{{\tt 1801.09605}}].

\bibitem{Grumiller:2002nm}
D.~Grumiller, W.~Kummer and D.~V. Vassilevich, {\it {Dilaton gravity in two-dimensions}},  {\em Phys. Rept.} {\bf 369} 327--430 [\href{http://arXiv.org/abs/hep-th/0204253}{{\tt hep-th/0204253}}].

\bibitem{Christensen:2013lma}
M.~H. Christensen, J.~Hartong, N.~A. Obers and B.~Rollier, {\it {Torsional Newton-Cartan Geometry and Lifshitz Holography}},  {\em Phys. Rev. D} {\bf 89} (2014) 061901 [\href{http://arXiv.org/abs/1311.4794}{{\tt 1311.4794}}].

\bibitem{Hartong:2015xda}
J.~Hartong, {\it {Gauging the Carroll Algebra and Ultra-Relativistic Gravity}},  {\em JHEP} {\bf 08} (2015) 069 [\href{http://arXiv.org/abs/1505.05011}{{\tt 1505.05011}}].

\bibitem{Hansen:2020pqs}
D.~Hansen, J.~Hartong and N.~A. Obers, {\it {Non-Relativistic Gravity and its Coupling to Matter}},  {\em JHEP} {\bf 06} (2020) 145 [\href{http://arXiv.org/abs/2001.10277}{{\tt 2001.10277}}].

\bibitem{Hansen:2021fxi}
D.~Hansen, N.~A. Obers, G.~Oling and B.~T. S{\o}gaard, {\it {Carroll Expansion of General Relativity}},  {\em SciPost Phys.} {\bf 13} (2022), no.~3 055 [\href{http://arXiv.org/abs/2112.12684}{{\tt 2112.12684}}].

\bibitem{Bergshoeff:2022eog}
E.~Bergshoeff, J.~Figueroa-O'Farrill and J.~Gomis, {\it {A non-lorentzian primer}},  {\em SciPost Phys. Lect. Notes} {\bf 69} (2023) 1 [\href{http://arXiv.org/abs/2206.12177}{{\tt 2206.12177}}].

\bibitem{Gomis:2000bd}
J.~Gomis and H.~Ooguri, {\it {Nonrelativistic closed string theory}},  {\em J. Math. Phys.} {\bf 42} (2001) 3127--3151 [\href{http://arXiv.org/abs/hep-th/0009181}{{\tt hep-th/0009181}}].

\bibitem{Danielsson:2000gi}
U.~H. Danielsson, A.~Guijosa and M.~Kruczenski, {\it {IIA/B, wound and wrapped}},  {\em JHEP} {\bf 10} (2000) 020 [\href{http://arXiv.org/abs/hep-th/0009182}{{\tt hep-th/0009182}}].

\bibitem{Cardona:2016ytk}
B.~Cardona, J.~Gomis and J.~M. Pons, {\it {Dynamics of Carroll Strings}},  {\em JHEP} {\bf 07} (2016) 050 [\href{http://arXiv.org/abs/1605.05483}{{\tt 1605.05483}}].

\bibitem{Bidussi:2021ujm}
L.~Bidussi, T.~Harmark, J.~Hartong, N.~A. Obers and G.~Oling, {\it {Torsional string Newton-Cartan geometry for non-relativistic strings}},  {\em JHEP} {\bf 02} (2022) 116 [\href{http://arXiv.org/abs/2107.00642}{{\tt 2107.00642}}].

\bibitem{Oling:2022fft}
G.~Oling and Z.~Yan, {\it {Aspects of Nonrelativistic Strings}},  {\em Front. in Phys.} {\bf 10} (2022) 832271 [\href{http://arXiv.org/abs/2202.12698}{{\tt 2202.12698}}].

\bibitem{Harksen:2024bnh}
M.~Harksen, D.~Hidalgo, W.~Sybesma and L.~Thorlacius, {\it {Carroll strings with an extended symmetry algebra}},  {\em JHEP} {\bf 05} (2024) 206 [\href{http://arXiv.org/abs/2403.01984}{{\tt 2403.01984}}].

\bibitem{Taylor:2015glc}
M.~Taylor, {\it {Lifshitz holography}},  {\em Class. Quant. Grav.} {\bf 33} (2016), no.~3 033001 [\href{http://arXiv.org/abs/1512.03554}{{\tt 1512.03554}}].

\bibitem{Leblonde:1965}
J.-M. L{\'e}vy-Leblond, {\it Une nouvelle limite non-relativiste du groupe de poincar{\'e}},  {\em Annales De L Institut Henri Poincare-physique Theorique} {\bf 3} (1965) 1--12.

\bibitem{Brugues:2006yd}
J.~Brugues, J.~Gomis and K.~Kamimura, {\it {Newton-Hooke algebras, non-relativistic branes and generalized pp-wave metrics}},  {\em Phys. Rev. D} {\bf 73} (2006) 085011 [\href{http://arXiv.org/abs/hep-th/0603023}{{\tt hep-th/0603023}}].

\bibitem{Derome1972HookesSA}
J.~R. Derome and J.~G. Dubois, {\it Hooke’s symmetries and nonrelativistic cosmological kinematics. — i},  {\em Il Nuovo Cimento B (1971-1996)} {\bf 9} (1972) 351--376.

\bibitem{Aldrovandi:1998im}
R.~Aldrovandi, A.~L. Barbosa, L.~C.~B. Crispino and J.~G. Pereira, {\it {Non-Relativistic spacetimes with cosmological constant}},  {\em Class. Quant. Grav.} {\bf 16} (1999) 495--506 [\href{http://arXiv.org/abs/gr-qc/9801100}{{\tt gr-qc/9801100}}].

\bibitem{Gao:2001sr}
Y.-h. Gao, {\it {Symmetries, matrices, and de Sitter gravity}},  {\em Conf. Proc. C} {\bf 0208124} (2002) 271--310 [\href{http://arXiv.org/abs/hep-th/0107067}{{\tt hep-th/0107067}}].

\bibitem{Grumiller:2020elf}
D.~Grumiller, J.~Hartong, S.~Prohazka and J.~Salzer, {\it {Limits of JT gravity}},  {\em JHEP} {\bf 02} (2021) 134 [\href{http://arXiv.org/abs/2011.13870}{{\tt 2011.13870}}].

\bibitem{Gomis:2020wxp}
J.~Gomis, D.~Hidalgo and P.~Salgado-Rebolledo, {\it {Non-relativistic and Carrollian limits of Jackiw-Teitelboim gravity}},  {\em JHEP} {\bf 05} (2021) 162 [\href{http://arXiv.org/abs/2011.15053}{{\tt 2011.15053}}].

\bibitem{Mertens:2022irh}
T.~G. Mertens and G.~J. Turiaci, {\it {Solvable models of quantum black holes: a review on Jackiw{\textendash}Teitelboim gravity}},  {\em Living Rev. Rel.} {\bf 26} (2023), no.~1 4 [\href{http://arXiv.org/abs/2210.10846}{{\tt 2210.10846}}].

\bibitem{Bacry:1968zf}
H.~Bacry and J.~Levy-Leblond, {\it {Possible kinematics}},  {\em J. Math. Phys.} {\bf 9} (1968) 1605--1614.

\bibitem{Gibbons:2003rv}
G.~W. Gibbons and C.~E. Patricot, {\it {Newton-Hooke space-times, Hpp waves and the cosmological constant}},  {\em Class. Quant. Grav.} {\bf 20} (2003) 5225 [\href{http://arXiv.org/abs/hep-th/0308200}{{\tt hep-th/0308200}}].

\bibitem{Coussaert:1995zp}
O.~Coussaert, M.~Henneaux and P.~van Driel, {\it {The Asymptotic dynamics of three-dimensional Einstein gravity with a negative cosmological constant}},  {\em Class. Quant. Grav.} {\bf 12} (1995) 2961--2966 [\href{http://arXiv.org/abs/gr-qc/9506019}{{\tt gr-qc/9506019}}].

\bibitem{Ivanov:1975zq}
E.~A. Ivanov and V.~I. Ogievetsky, {\it {The Inverse Higgs Phenomenon in Nonlinear Realizations}},  {\em Teor. Mat. Fiz.} {\bf 25} (1975) 164--177.

\bibitem{Afshar:2019axx}
H.~Afshar, H.~A. Gonz{\'a}lez, D.~Grumiller and D.~Vassilevich, {\it {Flat space holography and the complex Sachdev-Ye-Kitaev model}},  {\em Phys. Rev. D} {\bf 101} (2020), no.~8 086024 [\href{http://arXiv.org/abs/1911.05739}{{\tt 1911.05739}}].

\bibitem{Afshar:2019tvp}
H.~R. Afshar, {\it {Warped Schwarzian theory}},  {\em JHEP} {\bf 02} (2020) 126 [\href{http://arXiv.org/abs/1908.08089}{{\tt 1908.08089}}].

\bibitem{Kirillov:1976uta}
A.~A. Kirillov, {\em {Elements of the Theory of Representations}}, vol.~220 of {\em Grundlehren der mathematischen Wissenschaften}.
\newblock Springer Berlin, Heidelberg, 1~ed., 1976.

\bibitem{Kostant:1970qur}
B.~Kostant, {\em {\textit{Quantization and unitary representations}}}.
\newblock Springer, Berlin, Heidelberg, 1970.

\bibitem{Souriau:1970sdd}
J.-M. Souriau, {\em {\textit{Structure des systèmes dynamiques}}}.
\newblock Dunod, Paris, 1970.
\newblock Réimprimé par les éditions Jacques Gabay, 2008.

\bibitem{Woodard:2015zca}
R.~P. Woodard, {\it {Ostrogradsky's theorem on Hamiltonian instability}},  {\em Scholarpedia} {\bf 10} (2015), no.~8 32243 [\href{http://arXiv.org/abs/1506.02210}{{\tt 1506.02210}}].

\bibitem{Maldacena:2019cbz}
J.~Maldacena, G.~J. Turiaci and Z.~Yang, {\it {Two dimensional Nearly de Sitter gravity}},  {\em JHEP} {\bf 01} (2021) 139 [\href{http://arXiv.org/abs/1904.01911}{{\tt 1904.01911}}].

\bibitem{Cotler:2019nbi}
J.~Cotler, K.~Jensen and A.~Maloney, {\it {Low-dimensional de Sitter quantum gravity}},  {\em JHEP} {\bf 06} (2020) 048 [\href{http://arXiv.org/abs/1905.03780}{{\tt 1905.03780}}].

\bibitem{Stanford:2017thb}
D.~Stanford and E.~Witten, {\it {Fermionic Localization of the Schwarzian Theory}},  {\em JHEP} {\bf 10} (2017) 008 [\href{http://arXiv.org/abs/1703.04612}{{\tt 1703.04612}}].

\bibitem{Cartan:1923}
E.~Cartan, {\it {Sur les vari\'et\'es \`a connexion affine et la th\'eorie de la relativit\'e g\'en\'eralis\'ee (premi\`ere partie)}},  {\em Ann. \'Ec. Norm. Sup\'er.} {\bf 40} (1923) 325--412.

\bibitem{Cartan:1924}
E.~Cartan, {\it {Sur les vari\'et\'es \`a connexion affine et la th\'eorie de la relativit\'e g\'en\'eralis\'ee (suite)}},  {\em Ann. \'Ec. Norm. Sup\'er.} {\bf 41} (1924) 1--25.

\bibitem{Hartong:2022lsy}
J.~Hartong, N.~A. Obers and G.~Oling, {\it {Review on Non-Relativistic Gravity}},  {\em Front. in Phys.} {\bf 11} (2023) 1116888 [\href{http://arXiv.org/abs/2212.11309}{{\tt 2212.11309}}].

\bibitem{Hansen:2019vqf}
D.~Hansen, J.~Hartong and N.~A. Obers, {\it {Gravity between Newton and Einstein}},  {\em Int. J. Mod. Phys. D} {\bf 28} (2019), no.~14 1944010 [\href{http://arXiv.org/abs/1904.05706}{{\tt 1904.05706}}].

\bibitem{Bergshoeff:2014uea}
E.~A. Bergshoeff, J.~Hartong and J.~Rosseel, {\it {Torsional Newton{\textendash}Cartan geometry and the Schr{\"o}dinger algebra}},  {\em Class. Quant. Grav.} {\bf 32} (2015), no.~13 135017 [\href{http://arXiv.org/abs/1409.5555}{{\tt 1409.5555}}].

\bibitem{Hartong:2014oma}
J.~Hartong, E.~Kiritsis and N.~A. Obers, {\it {Lifshitz space{\textendash}times for Schr{\"o}dinger holography}},  {\em Phys. Lett. B} {\bf 746} (2015) 318--324 [\href{http://arXiv.org/abs/1409.1519}{{\tt 1409.1519}}].

\bibitem{Duval:1984cj}
C.~Duval, G.~Burdet, H.~P. Kunzle and M.~Perrin, {\it {Bargmann Structures and Newton-cartan Theory}},  {\em Phys. Rev. D} {\bf 31} (1985) 1841--1853.

\bibitem{Andringa:2016sot}
R.~Andringa, {\em {Newton-Cartan gravity revisited}}.
\newblock PhD thesis, High-Energy Frontier, Groningen U., Groningen U., 2016.

\bibitem{Blacker:2025zca}
M.~J. Blacker, A.~Castro, W.~Sybesma and C.~Toldo, {\it {Quantum corrections to the path integral of near extremal de Sitter black holes}},  {\em JHEP} {\bf 08} (2025) 120 [\href{http://arXiv.org/abs/2503.14623}{{\tt 2503.14623}}].

\bibitem{Maulik:2025phe}
S.~Maulik, A.~Mitra, D.~Mukherjee and A.~Ray, {\it {Logarithmic corrections to near-extremal entropy of charged de Sitter black holes}},  {\em JHEP} {\bf 01} (2026) 156 [\href{http://arXiv.org/abs/2503.08617}{{\tt 2503.08617}}].

\bibitem{Maulik:2024dwq}
S.~Maulik, L.~A. Pando~Zayas, A.~Ray and J.~Zhang, {\it {Universality in logarithmic temperature corrections to near-extremal rotating black hole thermodynamics in various dimensions}},  {\em JHEP} {\bf 06} (2024) 034 [\href{http://arXiv.org/abs/2401.16507}{{\tt 2401.16507}}].

\bibitem{PandoZayas:2026vbg}
L.~A. Pando~Zayas and J.~Zhang, {\it {A Universality Theorem for the Quantum Thermodynamics of Near-Extremal Black Holes}},  \href{http://arXiv.org/abs/2602.16767}{{\tt 2602.16767}}.

\bibitem{Witten:1987ty}
E.~Witten, {\it {Coadjoint Orbits of the Virasoro Group}},  {\em Commun. Math. Phys.} {\bf 114} (1988) 1.

\bibitem{Alekseev:1988ce}
A.~Alekseev and S.~L. Shatashvili, {\it {Path Integral Quantization of the Coadjoint Orbits of the Virasoro Group and 2D Gravity}},  {\em Nucl. Phys. B} {\bf 323} (1989) 719--733.

\bibitem{Saad:2019lba}
P.~Saad, S.~H. Shenker and D.~Stanford, {\it {JT gravity as a matrix integral}},  \href{http://arXiv.org/abs/1903.11115}{{\tt 1903.11115}}.

\end{thebibliography}\endgroup

\end{document}